\documentclass[aps]{revtex4}
\def\@llsymbol#1{\ifcase#1\or {}\or {'}\or {''}\or {'''}\or
   {''''}\or {'''''}\or  \else\@ctrerr\fi\relaz}
\newcounter{contador}
\newcommand{\letra}{
   \stepcounter{equation}
   \setcounter{contador}{\value{equation}}
   \setcounter{equation}{0}
   \renewcommand{\theequation}{\thecontador\alph{equation}}}
\newcommand{\antiletra}{
   \renewcommand{\theequation}{\arabic{equation}}
   \setcounter{equation}{\value{contador}}}

\usepackage{color}
\usepackage{indentfirst}
\usepackage{eucal}
\usepackage{amsmath}
\usepackage{amsfonts}
\usepackage{amssymb}
\usepackage{mathrsfs}
\begin{document}


%

%


\title{New solutions to the confluent Heun equation and quasiexact solvability}
\author{L\'ea Jaccoud El-Jaick}
\email{leajj@cbpf.br}
\author{Bartolomeu D. B. Figueiredo}
\email{barto@cbpf.br}
\affiliation{Centro Brasileiro de Pesquisas F\'{\i}sicas (CBPF),\\
Rua Dr. Xavier Sigaud, 150, CEP 22290-180, Rio de Janeiro, RJ, Brasil}
\begin{abstract}
\noindent
{\bf Abstract: }{We construct new solutions in series of 
confluent hypergeometric functions for
the confluent Heun equation (CHE).  Some of these solutions are
applied to the one-dimensional stationary Schr\"{o}dinger equation with hyperbolic
and trigonometric quasiexactly solvable potentials.}

\tableofcontents

\end{abstract}

\maketitle



%
\section{Introduction} 


By means of an integral
transformation, recently \cite{arxiv} we have found a new solution for 
the spheroidal wave equation and pointed out that such 
solution can generate a new group of solutions for the 
confluent Heun equation (CHE) -- this equation is more general than 
the spheroidal one. Now we establish explicitly that 
group of solutions and discuss, as illustration, possible applications 
to the one-dimensional stationary Schr\"{o}dinger equation with
quasiexactly solvable (QES) potentials.

For the CHE, or generalized spheroidal wave equation \cite{wilson},
we use the form \cite{leaver} 
\begin{eqnarray}
\label{gswe}
\displaystyle z(z-z_{0})\frac{d^{2}U}{dz^{2}}+(B_{1}+B_{2}z)\frac{dU}{dz}
+\displaystyle\left[B_{3}-2\omega\eta(z-z_{0})+\omega^{2}z(z-z_{0})\right]U=0,
\end{eqnarray}
where $z_{0}$, $B_{i}$, $\eta$ and $\omega$ are constants. 
In applications, we consider a particular case known 
as Whittaker-Hill equation (WHE) or Hill's equation with three terms. It is 
written in the form \cite{arscott,arscott3}
 %
 \begin{eqnarray}\label{whe}
 \frac{d^2W}{du^2}+\varsigma^2\left[\vartheta-\frac{1}{8}\xi^{2}
 -(p+1)\xi\cos(2\varsigma u)+
 \frac{1}{8}\xi^{2}\cos(4\varsigma u)\right]W=0, \quad (\varsigma=1,\,i)
 \end{eqnarray}
where $\vartheta$, $\xi$ and $p$ are parameters;
if $u$ is a real variable, this represents the WHE
when $\varsigma=1$ and the modified WHE when $\varsigma=i$.
The substitutions
\letra
\begin{eqnarray}\label{wheasgswe-1}
\begin{array}{l}
W(u)=U(z), \quad  z=\cos^{2}(\varsigma u),\qquad  [\varsigma=1,i]
\end{array}
\end{eqnarray}
transform the WHE (\ref{whe}) into the CHE (\ref{gswe}) with
\begin{eqnarray}\label{wheasgswe-2}
\begin{array}{l} 
z_{0}=1,\qquad
B_{1}=-\frac{1}{2}, \qquad B_{2}=1, \qquad
B_{3}=\frac{(p+1)\xi-\vartheta}{4}, \qquad i\omega=\frac{\xi}{2},
\qquad i\eta=\frac{p+1}{2}.
\end{array}
\end{eqnarray}
The (ordinary) spheroidal wave equation \cite{nist},
given in Appendix B, is another case of CHE 
which has only three parameters as the WHE.

We consider solutions for the CHE given by series whose 
coefficients satisfy three-term recurrence relations. Regarding the 
range of values assumed by the summation index $n$, we distinguish 
three types of series: 
two-sided infinite series if $n$ runs from minus to plus infinity, 
one-sided infinite series if $n\geq 0$, and finite series if 
$n$ has a lower and an upper limit. 
Two-sided infinite series are necessary to assure the 
convergence of solutions for equations in which 
there is no free parameter. 
However, when we truncate the series on the left-hand side 
by requiring that $n\geq 0$, 
the two-sided series give one-sided infinite series
which are suitable for equations with a free parameter; 
in turn, one-sided infinite series become finite series
for special values of the equation parameters (Appendix A).

In section II we study sets of one-sided expansions in series 
of confluent hypergeometric functions. Each set is constituted by three solutions, 
one in series of irregular confluent hypergeometric functions, and two in series
of regular functions: in general the former converges near the infinity,
whereas the others converge for finite values of $z$. 
In section III we show how such 
solutions can be transformed into two-sided series; these 
must satisfy the ratio test for $n\to\infty$ and for $n\to-\infty$, 
a fact which restricts the regions of convergence of some
solutions (in comparison with the regions of one-sided series).
{In both cases, the solutions are valid under 
some constraints on the parameters of the CHE.  
Consequences of these constraints appear in sections IV and V
where we aplly} the three types of series  
to the stationary Schr\"{o}dinger
equation with quasiexactly solvable (QES) potentials.
{Finally, in section VI we present
some conclusions and mention issues 
which deserve further investigations.}

{We recall that} for QES quantum-mechanical problems one part of 
energy spectrum and the respective eigenfunctions
can be computed explicitly \cite{turbiner,ushveridze1}.
If a QES problem obeys an equation of the
Heun family \cite{lea-1}, that part of
the spectrum may be derived from finite-series 
solutions if these are known. Indeed, a problem 
is QES if it admits solutions given by finite series whose coefficients
necessarily satisfy three-term or higher order
recurrence relations, and the problem 
is exactly solvable if admits 
solutions given by hypergeometric functions \cite{kalnins}.
This definition opens the possibility of finding the remaining part of the 
spectrum from one- or two-sided infinite-series solutions for 
the Heun equations.

We write the one-dimensional Schr\"{o}dinger
equation for a particle with mass $M$
and energy $E$ as 
\antiletra
  \begin{eqnarray}
  \label{schr}
  \frac{d^2\psi}{du^2}+\big[{\cal E}-\mathcal{V}(u)\big]\psi=0, 
  \quad u={a} x, \
   \quad {\cal E}=\frac{2M }{\hbar^2 a^2} E,
   \qquad \mathcal{V}(u)=\frac{2M }{\hbar^2 a^2} \mathrm{V}(x),
  \end{eqnarray}
 where ${a}$ is a constant with inverse-length dimension, 
 $\hbar$ is the Plank constant
 divided by $2\pi$, $x$ is the spatial coordinate and 
 ${V}(x)$ is the potential. 
 For $\mathcal{V}(u)$, in section IV we choose the 
  Ushveridze potential \cite{ushveridze1}
  \begin{eqnarray}
  \label{potencial}
  \mathcal{V}(u)&=&\begin{array}{l}4\beta^{2}{\sinh^{4}u}+ 4\beta\big[\beta-
  2(\gamma+\delta)-2\ell \big]{\sinh^{2}u}+4\left[ \delta-\frac{1}{4}\right]
   \left[\delta-\frac{3}{4}\right]\frac{1}{\sinh^{2}u}
   \end{array}
  \nonumber\vspace{2mm}\\
 &-& \begin{array}{l}
  4\left[ \gamma-\frac{1}{4}\right] \left[\gamma-\frac{3}{4}\right]\frac{1}
  {\cosh^{2}u},
  \end{array}
   \end{eqnarray}
where $\beta$, $\gamma$ and $\delta$ are real constants
with $\beta> 0$ and $\delta\geq 1/4$. Ushveridze stated that 
this potential is QES if $\ell=0,1,2,\cdots$.
However, we have found \cite{lea-2013} that, for
\begin{eqnarray}\label{4casos}
\begin{array}{l}
\left(\gamma,\delta\right)=\left(\frac{1}{4},\frac{1}{4}\right), \ 
\left(\frac{1}{4},\frac{3}{4}\right), \ 
\left(\frac{3}{4},\frac{1}{4}\right), \ 
\left(\frac{3}{4},\frac{3}{4}\right),
\end{array}
\end{eqnarray}
the potential (\ref{potencial}) is QES also when $\ell$ is a positive half-integer. 
{The above cases will be called Razavy-type potential because they
have the same form as a hyperbolic potential found by Razavy \cite{razavy} 
corresponding to a Schr\"{o}dinger equation which reduces to the modified WHE (\ref{whe}).}

For these Razavy-type potentials (\ref{4casos}) there are 
even and odd finite-series solutions 
which are bounded for any value of the independent variable. These afford  
a finite number of energy levels characteristic of QES problems. 
{Recently we have also found even infinite-series solutions 
\cite{lea-2013}, valid for any admissible value of $z$. We can get odd infinite-series 
solutions only if we use two solutions covering different intervals of $z$. 
We will see that the new expansions do 
not modify such results. However, by excluding the cases (\ref{4casos}), 
in addition to finite-series 
solutions we have already found infinite-series solutions 
bounded and convergent for $1/4\leq\delta<1/2$ and $1/2<\delta\leq 3/4$,  
without restrictions on the parameter $\gamma$ \cite{lea-2013}.
Now we will see that the new solutions
lead to a result less general in the sense that 
infinite-series solutions for $1/2<\delta\leq 3/4$ impose restrictions
on $\gamma$.}

In the section V, we briefly consider the Ushveridze trigonometric potential \cite{ushveridze1} 
\begin{eqnarray}\label{ush-trigonometrico}
\mathcal{V}(u)&=&\begin{array}{l}
-4\beta^{2}{\sin^{4}u}+ 4\beta\big[\beta+
2(\gamma+\delta+\ell) \big]{\sin^{2}u}+4\left[ \delta-\frac{1}{4}\right]
\left[\delta-\frac{3}{4}\right]\frac{1}{\sin^{2}u}
\end{array}
\nonumber\vspace{2mm}\\
&+& \begin{array}{l}
4\left[ \gamma-\frac{1}{4}\right] \left[\gamma-\frac{3}{4}\right]\frac{1}
{\cos^{2}u},
\end{array}\end{eqnarray}
 where $\beta$, $\gamma$ and $\delta$ are real constants
 with $\beta> 0$, $\gamma\geq 1/4$ and $\delta\geq 1/4$. As in the
 previous case, if $\ell=0,1,2,\cdots$ we have a QES problem ruled
 by the CHE. If $\gamma$ and $\delta$ are given by (\ref{4casos}), 
 once more the quasisolvability takes place also when $\ell$ is 
 a positive half-integer. Although the solutions in series of confluent 
 hypergeometric functions {and Coulomb wave functions
 may be used to deal with this problem,
 we conclude that we can simply use the Baber-Hass\'e
 expansions in power series \cite{lea-1,baber}.}

\section{One-sided series solutions}

First we present the confluent hypergeometric 
equation, remind the transformations
of the CHE and establish some notations. Thence we write
the expansions in series of confluent hypergeometric 
functions and notice that, in each set, one expansion has 
properties similar to the Baber-Hass\'e series.
After this, we show how the first set is constructed
and regard the convergence of the solutions.
\subsection{Confluent hypergeometric functions and symmetries
 of the CHE}

First, the regular and irregular confluent hypergeometric 
functions used as bases for the solutions of the CHE 
are denoted by $\Phi({a,c};u)$
and $\Psi({a,c};u)$, respectively. They satisfy the
confluent hypergeometric equation \cite{erdelii}
\begin{eqnarray}\label{confluent0}
u\frac{d^2\varphi(u)}{du^2}+\left( {c}-u\right)
\frac{d\varphi(u)}{du} -{a}\varphi(u)=0
\end{eqnarray}
which admits the solutions
\begin{eqnarray}\label{confluent1}
\begin{array}{ll}
\varphi^{1}(u)=\Phi({a,c};u),\qquad &
 \varphi^{2}(u)=u^{1-c}\Phi(1+a-c,2-c;u),\vspace{2mm}\\
%
\varphi^{3}(u)=\Psi(a,c;u),\quad &
 \varphi^{4}(u)=e^{u}u^{1-c}\Psi(1-a,2-c;-u),
 \end{array}
\end{eqnarray}
%
Alternative forms for these solutions
follow from the relations
\begin{eqnarray}\label{kummer}
\Phi({a,c};u)=e^{u}\Phi({c-a,c};-u),\qquad
\Psi({a,c};u)=u^{{1-c}}
\Psi(1+{a-c,2-c};u).
\end{eqnarray}
We take $\varphi^1$, $\varphi^2$ and
$\varphi^3$ as bases for the series expansions for the
solutions of the CHE: in this manner we form a set
of three expansions, one in series of irregular functions and two
in series of regular functions. If $c$ is not integer, in a common region 
of validity these three functions are linearly dependent
since
\begin{eqnarray}\label{continuacao}
\Psi(a,c;u)=\frac{\Gamma(1-c)}{\Gamma(a+1-c)}\Phi(a,c;u)+
\frac{\Gamma(c-1)}{\Gamma(a)}u^{1-c}\Phi(a+1-c,2-c;u)
\end{eqnarray}

Second, if $U(z)=U(B_{1},B_{2},B_{3}; z_{0},\omega,\eta;z)$
denotes one solution of the CHE, the following four transformations
$T_i$ leave invariant the form of the 
CHE \cite{lea-1,decarreau1,decarreau2}: 
%
\begin{eqnarray}\label{transformacao1}
\begin{array}{l}
T_{1}U(z)=z^{1+\frac{B_1}{z_0}}
U(C_{1},C_{2},C_{3};z_{0},\omega,\eta;z),\\
%
T_{2}U(z)=(z-z_{0})^{1-B_{2}-\frac{B_1}{z_0}}\ U(B_{1},D_{2},D_{3};
z_{0},\omega,\eta;z),\\
%
T_{3}U(z)=U(B_{1},B_{2},B_{3}; z_{0},-\omega,-\eta;z),
\\
%
T_{4}U(z)
=U(-B_{1}-B_{2}z_{0},B_{2},
B_{3}+2\eta\omega z_{0};z_{0},-\omega,
\eta;z_{0}-z),
\end{array}
\end{eqnarray}
where
%
\begin{eqnarray}\label{constantes-C-D}
&&\begin{array}{l}
C_{1}=-B_{1}-2z_{0}, \qquad
C_{2}=2+B_{2}+\frac{2B_{1}}{z_{0}},
\qquad C_{3}=B_{3}+
\left[1+\frac{B_{1}}{z_{0}}\right]
\left[B_{2}+\frac{B_{1}}{z_{0}}\right], \end{array}\nonumber\\
&&\begin{array}{l}
D_{2}=2-B_{2}-\frac{2B_{1}}{z_{0}},\qquad
D_{3}=B_{3}+
\frac{B_{1}}{z_{0}}\left(\frac{B_{1}}{z_{0}}
+B_{2}-1\right).\end{array}
\end{eqnarray}
Using the transformations (\ref{transformacao1}), 
from an initial set of three solutions  
it is possible to obtain a group constituted by
16 sets. For one-sided series solutions,
these sets of one-sided series solutions are denoted by
\begin{eqnarray}\label{notation-2}
\mathbb{\mathring{U}}_i(z)=\left[\mathring{U}_{i}^{\infty}(z),
\mathring{U}_{i}(z),
\mathring{\textsf{U}}_{i}(z)\right],
\end{eqnarray}
where the solutions $\mathring{U}_{ i}^{\infty}$ are given by series
of irregular confluent hypergeometric functions, 
whereas the solutions $\mathring{U}_{i}$ and 
$\mathring{\textsf{U}}_{i}$ are given by series of regular 
confluent hypergeometric functions. The ring 
over the symbols ($\mathring{U}$, for example)  
means that we are dealing with one-sided series solutions
($n\geq 0$). The infinity symbol appearing in 
$\mathring{U}_{i}^{\infty}(z)$ indicates that the series
associated with such solutions are possibly convergent
at $z=\infty$. 
We write only the 8 sets, namely,
\begin{eqnarray}\label{TRANS}
\begin{array}{llll}
\mathbb{\mathring{U}}_1(z),\qquad&
\mathbb{\mathring{U}}_2(z)=T_1\mathbb{\mathring{U}}_1(z),\quad&
\mathbb{\mathring{U}}_3(z)=T_2\mathbb{\mathring{U}}_2(z),\quad&
\mathbb{\mathring{U}}_4(z)=T_1\mathbb{\mathring{U}}_3(z);\vspace{2mm}\\
\mathbb{\mathring{U}}_5(z)=T_4\mathbb{\mathring{U}}_1(z),\quad&
\mathbb{\mathring{U}}_6(z)=T_4\mathbb{\mathring{U}}_2(z),&
\mathbb{\mathring{U}}_7(z)=T_4\mathbb{\mathring{U}}_3(z),&
\mathbb{\mathring{U}}_8(z)=T_4\mathbb{\mathring{U}}_4(z),
\end{array}
\end{eqnarray}
since others are obtained  by the transformation
$T_3$ which changes ($\eta,\omega$)
by ($-\eta,-\omega$) in the above sets.

In order to write the three solutions of a fixed set in terms of the same series 
coefficients $\mathring{b}_n^ i$ (where $n$ represents the summation
index), instead of ${\Phi}(a,b;y)$ we use the function 
$\tilde{\Phi}(a,b;y)$ defined by 
\begin{eqnarray}\label{phi-tilde}
\tilde{\Phi}(a,c;u)=\frac{\Gamma(c-a)}{\Gamma(c)}\ \Phi(a,c;u)=
\frac{\Gamma(c-a)}{\Gamma(c)}\left[1+\frac{a}{c}\frac{u}{1!}+\frac{a(a+1)}{c(c+1)}
\frac{u^2}{2!}+\cdots\right].
\end{eqnarray}
Thence, the series coefficients satisfy three-term recurrence relations 
having the form
\begin{eqnarray}\label{r1a}
\begin{array}{l}
\mathring{\alpha}_{0}^i \  \mathring{b}_{1}^i+
\mathring{\beta}_{0}^i\  \mathring{b}_{0}^i=0,\qquad
\mathring{\alpha}_{n}^i\  \mathring{b}_{n+1}^i+\mathring{\beta}_{n}^i\ 
\mathring{b}_{n}^i+
\mathring{\gamma}_{n}^i\  \mathring{b}_{n-1}^i=0\quad (n\geq1),
\end{array}
\end{eqnarray}
where $\mathring{\alpha}_{n}^i$, $\mathring{\beta}_{n}^i$ and 
$\mathring{\gamma} _{n}^i$
depend on the parameters of the differential equation.
If some of these parameters is arbitrary it may be determined
from the characteristic equation given by the infinite continued 
fraction \cite{leaver}
\begin{eqnarray}\label{transcendental}
\mathring{\beta}_{0}^i=\frac{\mathring{\alpha}_{0}^i
\mathring{\gamma}_{1}^i}{\mathring{\beta}_{1}^i-}\ 
\frac{\mathring{\alpha}_{1}^i
\mathring{\gamma}_{2}^i}
{\mathring{\beta}_{2}^i-}\ 
\frac{\mathring{\alpha}_{2}^i\mathring{\gamma}_{3}^i}{\mathring{\beta}_{3}^i-}\cdots ,
\end{eqnarray}
which is equivalent to the vanishing of the determinant 
of the tridiagonal matrix (\ref{matriz-tridiagonal}) associated with 
the previous relations.
\subsection{The sets of one-sided solutions and relations with power series}
In each of the following  sets, the solutions $\mathring{U}_{i}^{\infty}(z)$
and $\mathring{U}_{i}(z)$ are valid only if 
the parameters of the CHE satisfy certain restrictions which avoid that
the hypergeometric functions reduce to polynomials of fixed 
degree $l$. In effect, we have the relations \cite{erdelii}
\begin{eqnarray} \label{laguerre-1}
\tilde{\Phi}(-l,\alpha+1;u)=(-1)^l\Psi(-l,\alpha+1;u)=l!\ 
L_l^{(\alpha)}(u),\quad l=0,1,2,\cdots
\end{eqnarray}
where the $L_l^{(\alpha)}(u)$ denote generalized Laguerre polynomials
of degree $l$. 
Then, if $l$ is fixed (i. e., if $l$ does not depend
on the summation index $n$) $\tilde{\Phi}$ and $\Psi$
are polynomials of fixed degree and, so, cannot be
used as bases for series expansions. 
In addition, we have to
take into account that the functions
$\tilde{\Phi}(a,c;u)$ in general are well defined only if 
$c$ is not zero or negative integer.
Further, there are cases for which the definition
(\ref{phi-tilde}) is unsuitable because the argument of $\Gamma(c-a)$ is
zero or a negative integer: then, the solutions must be rewritten 
as in the case considered in Appendix B.
%

{\it First set:} ${2+i\eta-\frac{B_2}{2}
\neq 0,-1,\cdots}$ in $\mathring{U}^{\infty}_2(z)$ and $\mathring{U}_2(z)$.
\letra
\begin{equation}\label{che-primeiro-set}
\begin{array}{l}
\left[\begin{array}{l}
\mathring{U}_1^{\infty}(z)\vspace{2mm}\\
\mathring{U}_1(z)
\end{array}\right]=\begin{array}{l}e^{i\omega z}z^{1+\frac{B_1}{z_0}}
[z-z_0]^{1-B_2-\frac{B_1}{z_0}}\end{array}
\displaystyle\sum_{n=0}^{\infty}
\mathring{b}_n^{1}
\left[\begin{array}{l}
\Psi\left(2+i\eta-\frac{B_2}{2},2+\frac{B_1}{z_0}-n; - 2i\omega z\right)
\vspace{2mm}\\
\tilde{\Phi}\left(2+i\eta-\frac{B_2}{2},2+\frac{B_1}{z_0}-n; - 2i\omega z\right)
\end{array}\right],
\vspace{2mm}\\
\mathring{\textsf{U}}_1(z) = 
\begin{array}{l}
e^{i\omega z}
[z-z_0]^{1-B_2-\frac{B_1}{z_0}}\end{array}
\displaystyle\sum_{n=0}^{\infty}\begin{array}{l}
\mathring{b}_{n}^{1}\left(2i\omega z\right)^n \tilde{\Phi}
\left(n+1+i\eta-\frac{B_1}{z_0}-\frac{B_2}{2},
n-\frac{B_1}{z_0};-2i\omega z\right),\end{array}
\end{array}
\end{equation}
where series coefficients $\mathring{b}_n^1$ satisfy the recurrence relations 
(\ref{r1a}) with
\begin{eqnarray}\label{alfa}
&\begin{array}{l}
\mathring{\beta}_n^1=n\left[n+1-B_2-\frac{2B_1}{z_0}+
2i\omega z_0\right]+
\left[i\omega z_0-1-\frac{B_1}{z_0}\right] \left[2-B_2-
\frac{B_1}{z_0}\right]+2-B_2+B_3,\end{array}
\nonumber\\
&\begin{array}{l}
\mathring{\alpha}_n^1=- 2i\omega z_0(n+1), \qquad
\mathring{\gamma}_n^1=-\left[n+i\eta-\frac{B_1}{z_0}-\frac{B_2}{2}\right]
\left[n+1-B_2-\frac{B_1}{z_0}\right].\end{array}
\end{eqnarray}
This set is obtained in section II.B; in \cite{arxiv} it has been
written without further details.
%

{\it Second set:} ${1+i\eta-\frac{B_2}{2}-\frac{B_1}{z_0}
\neq 0,-1,\cdots}$ in $\mathring{U}^{\infty}_2(z)$ and $\mathring{U}_2(z)$.
\antiletra\letra
\begin{equation}\label{che-segundo-set}
\begin{array}{l}
\left[\begin{array}{l}
\mathring{U}^{\infty}_2(z)\vspace{2mm}\\
\mathring{U}_2(z)
\end{array}\right]=\begin{array}{l}e^{i\omega z}[z-z_0]^{1-B_2-\frac{B_1}{z_0}}\end{array}
\displaystyle\sum_{n=0}^{\infty}
\mathring{b}_n^{2}
\left[\begin{array}{l}
\Psi\left(1+i\eta-\frac{B_2}{2}-\frac{B_1}{z_0},-\frac{B_1}{z_0}-n; - 2i\omega z\right)
\vspace{2mm}\\
\tilde{\Phi}\left(1+i\eta-\frac{B_2}{2}-\frac{B_1}{z_0},-\frac{B_1}{z_0}-n; - 2i\omega z\right)
\end{array}\right],
\vspace{2mm}\\
\mathring{\textsf{U}}_2(z) = 
\begin{array}{l}
e^{i\omega z}z^{1+\frac{B_1}{z_0}}
[z-z_0]^{1-B_2-\frac{B_1}{z_0}}\end{array}
\displaystyle\sum_{n=0}^{\infty}\begin{array}{l}
\mathring{b}_{n}^{2}\left[2i\omega z\right]^n \tilde{\Phi}\left[n+2+i\eta-\frac{B_2}{2},
n+\frac{B_1}{z_0}+2;-2i\omega z\right];\end{array}
\end{array}
\end{equation}
\begin{eqnarray}\label{alfa-2}
&\begin{array}{l}
\mathring{\beta}_n^2=n\left[n+3-B_2+2i\omega z_0)\right]+
 i \omega z_0\left[2-B_2-\frac{B_1}{z_0}\right]+2-B_2+B_3, 
\end{array}
\nonumber\\
&\begin{array}{l}
\mathring{\alpha}_n^2=-2i\omega z_0(n+1), \qquad
\mathring{\gamma}_n^2=-\left[n+i\eta+1-\frac{B_2}{2}\right]
\left[n+1-B_2-\frac{B_1}{z_0}\right].\end{array}
\end{eqnarray}
%
%
%

{\it Third set:} $i\eta+\frac{B_2}{2}\neq 0,-1,\cdots$ in 
$\mathring{U}^{\infty}_3(z)$ and $\mathring{U}_3(z)$.
\antiletra\letra
\begin{equation}\label{che-third-set}
\begin{array}{l}
\left[\begin{array}{l}
\mathring{U}^{\infty}_3(z)\vspace{2mm}\\
\mathring{U}_3(z)
\end{array}\right]=\begin{array}{l}e^{i\omega z}\end{array}
\displaystyle\sum_{n=0}^{\infty}
\mathring{b}_n^{3}
\left[\begin{array}{l}
\Psi\left(i\eta+\frac{B_2}{2},-\frac{B_1}{z_0}-n; - 2i\omega z\right)
\vspace{2mm}\\
\tilde{\Phi}\left(i\eta+\frac{B_2}{2},-\frac{B_1}{z_0}-n; - 2i\omega z\right)
\end{array}\right],
\vspace{2mm}\\
\mathring{\textsf{U}}_3(z) = 
\begin{array}{l}
e^{i\omega z}z^{1+\frac{B_1}{z_0}}
\end{array}
\displaystyle\sum_{n=0}^{\infty}\begin{array}{l}
\mathring{b}_{n}^{3}\left[2i\omega z\right]^n \tilde{\Phi}\left[n+1+i\eta+\frac{B_2}{2}+\frac{B_1}{z_0},
n+\frac{B_1}{z_0}+2;-2i\omega z\right];\end{array}
\end{array}
\end{equation}
\begin{eqnarray}\label{alfa-3}
&\begin{array}{l}
\mathring{\beta}_n^3=n\left[n+1+B_2+
\frac{2B_1}{z_0}+2i\omega z_0\right]+
\left[B_2+\frac{B_1}{z_0}\right]\left[1+\frac{B_1}{z_0}+i\omega z_0\right]+B_3, 
\end{array}
\nonumber\\
&\begin{array}{l}
\mathring{\alpha}_n^3= -2i\omega z_0(n+1), \qquad
\mathring{\gamma}_n^3=-\left[n+i\eta+\frac{B_2}{2}+\frac{B_1}{z_0}\right]
\left[n-1+B_2+\frac{B_1}{z_0}\right].\end{array}
\end{eqnarray}
%
%
\antiletra
%

{\it Fourth set:} $i\eta+1+\frac{B_2}{2}+\frac{B_1}{z_0}\neq 0,-1,\cdots$
in $\mathring{U}^{\infty}_4(z)$ and $\mathring{U}_4(z)$.
\letra
\begin{equation}\label{che-fourth-set}
\begin{array}{l}
\left[\begin{array}{l}
\mathring{U}^{\infty}_4(z)\vspace{2mm}\\
\mathring{U}_4(z)
\end{array}\right]=\begin{array}{l}e^{i\omega z}z^{1+\frac{B_1}{z_0}}\end{array}
\displaystyle\sum_{n=0}^{\infty}
\mathring{b}_n^{4}
\left[\begin{array}{l}
\Psi\left(i\eta+1+\frac{B_2}{2}+\frac{B_1}{z_0} ,2+\frac{B_1}{z_0}-n; -2i\omega z\right)
\vspace{2mm}\\
\tilde{\Phi}\left(i\eta+1+\frac{B_2}{2}+\frac{B_1}{z_0} ,2+\frac{B_1}{z_0}-n; -2i\omega z\right)
\end{array}\right],
\vspace{2mm}\\
\mathring{\textsf{U}}_4(z) = 
\begin{array}{l}
e^{i\omega z}
\end{array}
\displaystyle\sum_{n=0}^{\infty}\begin{array}{l}
\mathring{b}_{n}^{4}\left[2i\omega z\right]^n \tilde{\Phi}\left[n+i\eta+\frac{B_2}{2},
n-\frac{B_1}{z_0};-2i\omega z\right];\end{array}
\end{array}
\end{equation}
\begin{eqnarray}\label{alfa-4}
&\begin{array}{l}
\mathring{\beta}_n^4=n\left[n-1+B_2+2i\omega z_0\right]+i\omega z_0
\left[B_2+\frac{B_1}{z_0}\right]+B_3, 
\end{array}
\nonumber\\
&\begin{array}{l}
\mathring{\alpha}_n^4= -2i\omega z_0(n+1), \qquad
\mathring{\gamma}_n^4=-\left[n+i\eta-1+\frac{B_2}{2}\right]\left[n-1+B_2+\frac{B_1}{z_0}\right].\end{array}
\end{eqnarray}
%
%
%

{\it Fifth set:} $2+i\eta-\frac{B_2}{2}\neq 0,-1,\cdots$ 
in $\mathring{U}^{\infty}_5(z)$ and $\mathring{U}_5(z)$.
\antiletra\letra
\begin{equation}\label{che-fifth-set}
\begin{array}{l}
\left[\begin{array}{l}
\mathring{U}^{\infty}_5(z)\vspace{2mm}\\
\mathring{U}_5(z)
\end{array}\right]=\begin{array}{l}e^{i\omega z}\,f(z) \end{array}
\displaystyle\sum_{n=0}^{\infty}
\mathring{b}_n^{5}
\left[\begin{array}{l}
\Psi\left(2+i\eta-\frac{B_2}{2},2-B_2 -\frac{B_1}{z_0}-n; 2i\omega (z_0-z)\right)
\vspace{2mm}\\
\tilde{\Phi}\left(2+i\eta-\frac{B_2}{2},2-B_2 -\frac{B_1}{z_0}-n; 2i\omega (z_0-z)\right)
\end{array}\right],
\vspace{2mm}\\
\mathring{\textsf{U}}_5 = 
\begin{array}{l}
e^{i\omega z}z^{1+\frac{B_1}{z_0}}
\end{array}
\displaystyle\sum_{n=0}^{\infty}\begin{array}{l}
\mathring{b}_{n}^{5}\left[2i\omega (z-z_0)\right]^n \tilde{\Phi}\left[n+1+
i\eta+\frac{B_2}{2}+ \frac{B_1}{z_0},n+B_2+\frac{B_1}{z_0};2i\omega (z_0-z)\right]
\end{array}
\end{array}
\end{equation}
where $f(z)=z^{1+\frac{B_1}{z_0}} [z-z_0]^{1-B_2-\frac{B_1}{z_0}}$ and
\begin{eqnarray}\label{alfa-5}
\begin{array}{l}
\mathring{\beta}_n^5=n\left[n+1+B_2+\frac{2B_1}{z_0}-2i\omega z_0\right]+
\frac{B_1}{z_0}\left[B_2+\frac{B_1}{z_0}+1-i\omega z_0\right] + 
2\eta \omega z_0-2i\omega z_0  
\end{array}
\nonumber\\
\begin{array}{l}
+B_2+B_3,\qquad
\mathring{\alpha}_n^5= 2i\omega z_0(n+1), \qquad
\mathring{\gamma}_n^5=-\left[n+i\eta+\frac{B_2}{2}+\frac{B_1}{z_0}\right]
\left[n+1+\frac{B_1}{z_0}\right].\end{array}
\end{eqnarray}
%
%

%

{\it Sixth set:} $i\eta+1+\frac{B_1}{z_0}+\frac{B_2}{2} \neq 0,-1,
\cdots$ in $\mathring{U}^{\infty}_{6}(z)$ 
and $\mathring{U}_{6}(z)$.
\antiletra\letra
\begin{equation}\label{che-six-set}
\begin{array}{l}
\left[\begin{array}{l}
\mathring{U}^{\infty}_6\vspace{2mm}\\
\mathring{U}_6
\end{array}\right]=\begin{array}{l}e^{i\omega z}  z^{1+\frac{B_1}{z_0}} \end{array}
\displaystyle\sum_{n=0}^{\infty}
\mathring{b}_n^{6}
\left[\begin{array}{l}
\Psi\left[1+i\eta+\frac{B_1}{z_0}+\frac{B_2}{2},B_2 +\frac{B_1}{z_0}-n;
 2i\omega (z_0-z)\right]\vspace{2mm}\\
\tilde{\Phi}\left[1+i\eta+\frac{B_1}{z_0}+\frac{B_2}{2},B_2 +\frac{B_1}{z_0}-n;
 2i\omega (z_0-z)\right]\end{array}\right],
\vspace{2mm}\\
\mathring{\textsf{U}}_6 = 
\begin{array}{l}
e^{i\omega z}\end{array} f(z) 
\displaystyle\sum_{n=0}^{\infty}\begin{array}{l}
\mathring{b}_{n}^{6}\left[2i\omega (z-z_0)\right]^n \tilde{\Phi}\left[n+2+i\eta-\frac{B_2}{2},
n+2-B_2-\frac{B_1}{z_0};2i\omega (z_0-z)\right]\end{array}
\end{array}
\end{equation}
where $f(z)$ is given as in (\ref{che-fifth-set}), and
\begin{eqnarray}\label{alfa-6}
&\begin{array}{l}
\mathring{\beta}_n^6=n\left[n+3-2i\omega z_0-B_2\right]
-i\omega B_1+2\eta\omega z_0 
-2i\omega z_0+B_3 -B_2+2,
\end{array}
\nonumber\\
&\begin{array}{l}
\mathring{\alpha}_n^6= 2i\omega z_0(n+1), \qquad
\mathring{\gamma}_n^6=-\left[n+i\eta{+1-\frac{B_2}{2}}\right]
\left[n{+1+\frac{B_1}{z_0}}\right].\end{array}
\end{eqnarray}
%
%

{\it Seventh set:} $i\eta+\frac{B_2}{2}\neq 0,-1,\cdots$ in 
$\mathring{U}^{\infty}_{7}(z)$ 
and $\mathring{U}_{7}(z)$.
\antiletra\letra
\begin{equation}\label{che-seventh-set}
\begin{array}{l}
\left[\begin{array}{l}
\mathring{U}^{\infty}_7(z)\vspace{2mm}\\
\mathring{U}_7(z)
\end{array}\right]=\begin{array}{l}e^{i\omega z}  \end{array}
\displaystyle\sum_{n=0}^{\infty}
\mathring{b}_n^{7}
\left[\begin{array}{l}
\Psi\left(i\eta+\frac{B_2}{2},B_2 +\frac{B_1}{z_0}-n; 2i\omega (z_0-z)\right)
\vspace{2mm}\\
\tilde{\Phi}\left(i\eta+\frac{B_2}{2},B_2 +\frac{B_1}{z_0}-n; 2i\omega (z_0-z)\right)
\end{array}\right],
\vspace{2mm}\\
\mathring{\textsf{U}}_7 = 
\begin{array}{l}g(z)
\end{array} 
\displaystyle\sum_{n=0}^{\infty}\begin{array}{l}
\mathring{b}_{n}^{7}\left[2i\omega (z-z_0)\right]^n \tilde{\Phi}\left[n+1+i\eta-\frac{B_1}{z_0}-\frac{B_2}{2},
n+2-B_2-\frac{B_1}{z_0};2i\omega (z_0-z)\right]\end{array}
\end{array}
\end{equation}
where $g(z)=e^{i\omega z}[z-z_0]^{1-B_2-\frac{B_1}{z_0}}$ and
\begin{eqnarray}\label{alfa-7}
&\begin{array}{l}
\mathring{\beta}_n^7=n\left[n+1-2i\omega z_0-B_2-\frac{2B_1}{z_0}\right]+
i\omega  B_1 +2\eta\omega z_0 +B_3 + \frac{B_1}{z_0}\left[\frac{B_1}{z_0}+B_2-1\right],
\end{array}
\nonumber\\
&\begin{array}{l}
\mathring{\alpha}_n^7= 2i\omega z_0(n+1), \qquad
\mathring{\gamma}_n^7=-\left[n+i\eta{-\frac{B_1}{z_0}-\frac{B_2}{2}}\right]
\left[n{-1-\frac{B_1}{z_0}}\right].\end{array}
\end{eqnarray}
%
%

{\it Eighth set:} $i\eta+1-\frac{B_2}{2}-\frac{B_1}{z_0}\neq 0,-1,
\cdots$ in $\mathring{U}^{\infty}_8(z)$ and $\mathring{U}_8(z)$.
\antiletra\letra
\begin{equation}\label{che-eight-set}
\begin{array}{l}
\left[\begin{array}{l}
\mathring{U}^{\infty}_8(z)\vspace{2mm}\\
\mathring{U}_8(z)
\end{array}\right]=\begin{array}{l} \,g(z)     \end{array}
\displaystyle\sum_{n=0}^{\infty}
\mathring{b}_n^{8}
\left[\begin{array}{l}
\Psi\left[i\eta+1-\frac{B_2}{2}-\frac{B_1}{z_0},2-B_2 -\frac{B_1}{z_0}-n;
 2i\omega (z_0-z)\right]
\vspace{2mm}\\
\tilde{\Phi}\left[i\eta+1-\frac{B_2}{2}-\frac{B_1}{z_0},2-B_2 -\frac{B_1}{z_0}-n; 
2i\omega (z_0-z)\right]
\end{array}\right],
\vspace{2mm}\\
\mathring{\textsf{U}}_8(z) = 
\begin{array}{l}
e^{i\omega z}\end{array}
\displaystyle\sum_{n=0}^{\infty}\begin{array}{l}
\mathring{b}_{n}^{8}\left[2i\omega (z-z_0)\right]^n \tilde{\Phi}\left[n+i\eta+
\frac{B_2}{2},n+B_2+\frac{B_1}{z_0};2i\omega (z_0-z)\right]\end{array}
\end{array}
\end{equation}
where $g(z)$ is given in $\mathring{\textsf{U}}_7$  
and
\begin{eqnarray}\label{alfa-8}
&\begin{array}{l}
\mathring{\beta}_n^8=n\left[n-1+B_2-2i\omega z_0\right]+i\omega B_1 +2\eta\omega z_0 +B_3,
\end{array}
\nonumber\\
&\begin{array}{l}
\mathring{\alpha}_n^8= 2i\omega z_0(n+1), \qquad
\mathring{\gamma}_n^8=-\left[n-1+i\eta+\frac{B_2}{2}\right]
\left[n-1-\frac{B_1}{z_0}\right].\end{array}
\end{eqnarray}

In each set, the solution $\mathring{\textsf{U}}_i$
is the only one affording finite series which can be 
expressible as a sum of polynomials. 
In effect, if the parameter $a$ of 
$\tilde{\Phi}(a,c;y)$ in $\mathring{\textsf{U}}_i$
is $a=n-N$ (where $N=0,1,2,\cdots$), 
then $\mathring{\textsf{U}}_i$ is given by finite 
 series since $\mathring{\gamma}_n^i\propto(n-N-1)$, a fact
which implies series terminating at $n=N$ (see Appendix A). 
In this case, the functions $\tilde{\Phi}(n-N,c;y)$ reduce to the  
Laguerre polynomials (\ref{laguerre-1}) with $l=N-n$ ($0\leq n\leq N$). 
This fact allows us to obtain some relations among the solutions
$\mathring{\textsf{U}}_i(z)$ and the 
Baber-Hass\'e solutions in power series for the CHE
\cite{lea-1,baber}. We denote the latter 
by $U_i^{\text{baber}}(z)$ and take 
\antiletra\letra
\begin{eqnarray}\label{barber-1}
&U_{1}^{\text{baber}}(z)=e^{i\omega z}\displaystyle \sum_{n=0}^{\infty}a_{n}^{1}
(z-z_{0})^{n}\qquad \text{with}\\
%
&\begin{array}{l}
 z_{0}\left[n+B_{2}+\frac{B_{1}}{z_{0}}\right]
\left[n+1\right]a_{n+1}^{1}+
\mathring{\beta}_n^{4}\,a_{n}^{1}+
2i\omega\left[n+i\eta+\frac{B_{2}}{2}-1\right]
a_{n-1}^{1}=0,\quad a_{-1}^{1}=,0\quad
\end{array}
\end{eqnarray}
where $\mathring{\beta}_n^{4}$ is given in (\ref{alfa-4}).
Eight solutions are generated as in (\ref{TRANS}), that is,
\antiletra
\begin{eqnarray}\label{TRANS-2}
\begin{array}{llll}
{U}_1^{\text{baber}},\qquad&
{U}_2^{\text{baber}}=T_1U_1^{\text{baber}},\quad&
U_3^{\text{baber}}=T_2U_2^{\text{baber}},\quad&
U_4^{\text{baber}}=T_1U_3^{\text{baber}},\vspace{2mm}\\
U_5^{\text{baber}}=T_4U_1^{\text{baber}},\quad&
U_6^{\text{baber}}=T_4U_2^{\text{baber}},&
U_7^{\text{baber}}=T_4U_3^{\text{baber}},&
U_8^{\text{baber}}=T_4U_4^{\text{baber}},
\end{array}
\end{eqnarray}
while additional solutions are obtained by 
applying the transformation
$T_3$ on the above ones. The solution 
$U_5^{\text{baber}}$, to be used later, reads
\antiletra\letra
\begin{eqnarray}\label{barber-5}
&U_{5}^{\text{baber}}(z)=e^{i\omega z}\displaystyle \sum_{n=0}^{\infty}
(-)^na_{n}^{1}
z^{n}\qquad \text{with}\\
%
&\begin{array}{l}
 z_{0}\left[n-\frac{B_{1}}{z_{0}}\right]
\left[n+1\right]a_{n+1}^{5}+
\mathring{\beta}_n^{8}\,a_{n}^{5}-
2i\omega\left[n+i\eta+\frac{B_{2}}{2}-1\right]
a_{n-1}^{1}=0,\quad a_{-1}^{5}=0,\quad
\end{array}
\end{eqnarray}
where $\mathring{\beta}_n^{8}$ is given in (\ref{alfa-8}).
Notice that the coefficient
$\mathring{\beta}_n^{i}$ are the same of 
the solutions $\mathring{\textsf{U}}_i$.
In fact, the same $\mathring{\beta}_n^{i}$  
appears in the following pairs of solutions:
\antiletra
\begin{eqnarray}\label{hiper-baber}
\begin{array}{llll}
\mathring{\textsf{U}}_1\leftrightarrow U_4^{\text{baber}},\qquad&
\mathring{\textsf{U}}_2\leftrightarrow U_3^{\text{baber}},\quad&
\mathring{\textsf{U}}_3\leftrightarrow  U_2^{\text{baber}},\quad&
\mathring{\textsf{U}}_4\leftrightarrow  U_1^{\text{baber}},\vspace{2mm}\\
\mathring{\textsf{U}}_5\leftrightarrow  U_8^{\text{baber}},\quad&
\mathring{\textsf{U}}_6\leftrightarrow  U_7^{\text{baber}},&
\mathring{\textsf{U}}_7\leftrightarrow  U_6^{\text{baber}},&
\mathring{\textsf{U}}_8\leftrightarrow  U_5^{\text{baber}}.
\end{array}
\end{eqnarray}
Notice that the solutions $\mathring{\textsf{U}}_i$
in general are valid only if the parameter $c$ of
$\tilde{\Phi}(a,c;y)$ is not zero or negative integer,
a restriction not required by $U_i^{\text{baber}}$.
Nevertheless, if both solutions of a given pair are 
valid, we verify that

\begin{itemize}
\itemsep-3pt
\item
Both solutions in each pair (\ref{hiper-baber}) are given only 
by finite series or by infinite series.
\item
In each pair, the product $\mathring{\alpha}_n
\mathring{\gamma}_{n+1}$ is the same for both solutions
which, by this reason, present the same characteristic
equation having the form (\ref{transcendental}).
\item
As we will see, the infinite-series solutions $\mathring{\textsf{U}}_i$ 
converge for any finite $z$, just as the corresponding solutions $U_i^{\text{baber}}$.
\end{itemize}
For the Whittaker-Hill equation ($z_0=1$, $B_1=-1/2$, $B_2=1$), 
all the above pairs are valid, as we can check. In addition,
according to Appendix B, for the spheroidal equation we 
can take $z_0=1$, $B_2=-2B_1$ and $\eta=0$; hence, the
solutions $\mathring{\textsf{U}}_2$, $\mathring{\textsf{U}}_4$,
$\mathring{\textsf{U}}_6$ and $\mathring{\textsf{U}}_8$ 
reduce to Baber-Hass\'e expansions up to a redefinition
of the series coefficients.

\subsection{Construction of the first set of solutions}

It it necessary to study only the first set of solutions; 
the others follow from the transformations (\ref{transformacao1}).
Thus, to get the solutions (\ref{che-primeiro-set}), 
first we accomplish the substitutions
%
\begin{eqnarray}\label{Y}
U(z)=e^{i\omega z}z^{1+\frac{B_1}{z_0}}
(z-z_0)^{1-B_2-\frac{B_1}{z_0}}\ Y(y),
\qquad
y=-2i\omega z
\end{eqnarray}
which, when inserted into (\ref{gswe}), lead to
\begin{eqnarray}\label{V}
(y+2i\omega z_0)\left(y\frac{d^{2}Y}{dy^{2}}-
y\frac{dY}{dy}\right)+
\left(-2i \omega \bar{C}_1+\bar{C}_2y\right)\frac{dY}{dy}
\nonumber
\\
%
+\left[\bar{C}_{3}+i\omega \bar{C}_1+2\eta\omega z_0
-\frac{1}{2}(\bar{C}_2+2i\eta)y\right]Y=0,
\end{eqnarray}
where $\bar{C}_1=-B_1-2z_0,\ \bar{C}_2=4-B_2$ and 
$\bar{C}_3=B_3+2-B_2$.
%
%
In the second place, to obtain 
$\mathring{U}^{\infty}_1$ and $\mathring{U}_1$,
we expand $Y(y)$ as
\letra
\begin{eqnarray}\label{sph-2a}
Y(y)=
\sum_{n=0}^{\infty}\mathring{b}_n^{1}\mathscr{F}_n( y),
\qquad \mathscr{F}_n(y)=\Psi(\alpha,\beta-n;y) \text{ or } 
\mathscr{F}_n(y)=\tilde{\Phi}(\alpha,\beta-n;y)
\end{eqnarray}
and show that 
\begin{eqnarray}\label{sph-2a-}
\alpha=2+i\eta-(B_2/2),\qquad \beta=2+(B_1/z_0),
\end{eqnarray}
while the coefficients $\mathring{b}_n^1$ satisfy the relations 
(\ref{r1a}) with (\ref{alfa}). To this end, we use \cite{nist}
\antiletra
\begin{eqnarray}\label{using}
&&\begin{array}{l}
y\frac{d^2\mathscr{F}_n}{dy^2}-y\frac{d\mathscr{F}_n}{dy}=
(n-\beta)\frac{\mathscr{F}_n}{dy}+\alpha\mathscr{F}_n,
\qquad[\text{see Eq. (\ref{confluent0})}]\nonumber
\vspace{2mm}\\
\frac{d\mathscr{F}_n}{dy}=\mathscr{F}_n-\mathscr{F}_{n-1},
\end{array}\nonumber
\vspace{2mm}\\
&&
\begin{array}{l}y\mathscr{F}_{n-1}=
(n+1+\alpha-\beta)\mathscr{F}_{n+1}+(y+\beta-n-1)
\mathscr{F}_n.\end{array}
\end{eqnarray}
%
Inserting $Y(y)=\sum \mathring{b}_n^1\mathscr{F}_n(y)$ into
(\ref{V}) and using (\ref{using}), we find
\begin{eqnarray}\label{convergencia-1}
\displaystyle\sum_{n=0}^{\infty}\mathring{b}_n^{1}
 \mathring{\alpha}_{n-1}^{1}\mathscr{F}_{n-1}(y)
+\sum_{n=0}^{\infty}\mathring{b}_n^{1}\mathring{\beta}_{n}^{1} \mathscr{F}_{n}(y)+
\sum_{n=0}^{\infty} \mathring{b}_n^{1}\mathring{\gamma}_{n+1}^{1}\mathscr{F}_{n+1}(y)+
\nonumber\\
\displaystyle\sum_{n=0}^{\infty}
\begin{array}{l}
 \mathring{b}_n^{1}
\left[\alpha-i\eta-\frac{1}{2}\bar{C}_2\right]y\mathscr{F}_{n}(y)
=0\end{array}
\end{eqnarray}
or, equivalently, 
\begin{eqnarray}\label{r0}
&\mathring{\alpha}_{-1}^{1} \mathring{b}_{0}^{1} \mathscr{F}_{-1}(y)+
\left[\mathring{\alpha}_{0}^{1} \mathring{b}_{1}^{1}+
\mathring{\beta}_{0}^{1}\mathring{b}_{0}^{1}\right]\mathscr{F}_{0}+
\displaystyle\sum_{n=1}^{\infty}
\left[\mathring{\alpha}_{n}^{1} \mathring{b}_{n+1}^{1}+
\mathring{\beta}_{n}^{1} \mathring{b}_n^{1}+
\mathring{\gamma}_{n}^{1} \mathring{b}_{n-1}^{1}\right]
\mathscr{F}_{n}(y)
+
\nonumber\\
&\displaystyle\sum_{n=0}^{\infty} \begin{array}{l}
\mathring{b}_n^{1}
\left[\alpha-i\eta-\frac{1}{2}\bar{C}_2\right]y\mathscr{F}_{n}(y)
=0,\end{array}
\end{eqnarray}
where 
\begin{eqnarray*}
&&\begin{array}{l}
\mathring{\beta}_n^1=2i\omega z_0\left[n+\alpha-\beta\right]+
\left[n+1-\beta\right] \left[n-\beta+\bar{C}_2\right]+2\eta\omega 
z_0-i\omega\bar{C}_1+\bar{C}_3,\end{array}
\nonumber\\
&&\begin{array}{l}
\mathring{\alpha}_n^1=- 2i\omega z_0\left(n+1-\beta-\frac{\bar{C}_1}{z_0}\right), \qquad
\mathring{\gamma}_n^1=-\left[n+\alpha-\beta\right]
\left[n-1-\beta+\bar{C}_2\right].\end{array}
\end{eqnarray*}
Now we determine the constants $\alpha$ and $\beta$
by equating to zero the first and the last terms of Eq. (\ref{r0}).
More precisely, we take
\begin{eqnarray*}\begin{array}{l}
\mathring{\alpha}_{-1}^1=2i\omega z_0\left(\beta+\dfrac{1}{z_0}\bar{C}_1\right)=0,
\qquad  \alpha-i\eta-\frac{1}{2}\bar{C}_2=0,
\end{array}
\end{eqnarray*}
what demands that $\beta=2+(B_1/z_0)$ and $\alpha=2+i\eta-(B_2/2)$, 
as stated in (\ref{sph-2a-}). Thence, Eq. (\ref{r0}) is satisfied if
the $\mathring{b}_n^1$ obey the relations (\ref{r1a}) with 
the coefficients (\ref{alfa}).

On the other hand, by supposing that Eq. (\ref{continuacao})
is valid, we can obtain $\mathring{\textsf{U}}_1$ 
as a linear combination of $\mathring{U}_1$ 
and $\mathring{U}_1^{\infty}$, although the validity of $\mathring{\textsf{U}}_1$
does not depend on Eq. (\ref{continuacao}). Actually, 
up to a multiplicative constant, we have
\begin{eqnarray*}
\mathring{\textsf{U}}_1(z)& = &
\begin{array}{l}
e^{i\omega z}z^{1+\frac{B_1}{z_0}}
[z-z_0]^{1-B_2-\frac{B_1}{z_0}}\end{array}
\displaystyle\sum_{n=0}^{\infty}\Big[\begin{array}{l}(-1)^n
\mathring{b}_{n}^{1}\left(-2i\omega z\right)^{n-1-\frac{B_1}{z_0}}
\times\end{array}\vspace{2mm}
\nonumber\\
&&
\begin{array}{l}
 \tilde{\Phi}\left(n+1+i\eta-\frac{B_1}{z_0}-\frac{B_2}{2},
n-\frac{B_1}{z_0};-2i\omega z\right)\end{array}\Big].
\end{eqnarray*}
Thus, in Eq. (\ref{Y}) 
\letra
\begin{equation}
Y(y)=
\sum_{n=0}^{\infty}\begin{array}{l}
(-1)^ n\mathring{b}_{n}^{1} \,y^{\text{c}_n-1} 
\tilde{\Phi}\left(\text{a}_n,
\text{c}_n;y\right),\quad\text{a}_n=n+1+i\eta-\frac{B_1}{z_0}-\frac{B_2}{2},
\: \text{c}_n=n-\frac{B_1}{z_0}.
\end{array}
\end{equation}
If, by definition we take
\begin{eqnarray}\label{def}
F_n(y):=(-1)^ny^{\text{c}_n-1} 
\tilde{\Phi}\left(\text{a}_n, \text{c}_n;y\right)  \quad  \Rightarrow\quad 
Y(y)=\sum_{n=0}^{\infty}
\mathring{b}_{n}^{1} F_n(y),
\end{eqnarray}
then we find the relations
%
%
%
%
\antiletra
\begin{eqnarray}\label{convergencia-2}
&&\begin{array}{l}
y\frac{d^2F_n(y)}{dy^2}-y\frac{dF_n(y)}{dy}=
\left(\text{c}_n-2\right)\frac{dF_n(y)}{dy} +\left(\text{a}_n+1-\text{c}_n\right)F_n\left(y\right),
\end{array}\nonumber\\
&&\begin{array}{l}\frac{dF_n(y)}{dy}=F_n(y)-F_{n-1}(y),\end{array}\nonumber\\
&&\begin{array}{l} 
yF_{n-1}(y)=\text{a}_nF_{n+1}(y)-
\left(\text{c}_n-1-y\right)
F_n(y).\end{array}
\end{eqnarray}
The first equation expresses the fact that $F_n(y)$
is basically the solution $\varphi^2(u)$ given in Eq. 
(\ref{confluent1}) for equation 
(\ref{confluent0}) with $u=y$, $a=\text{a}_n+1-\text{c}_n$ and $c=2-\text{c}_n$.
The second and third relations result, for example, from 
Eqs. (13.3.21) and (13.3.14) of Ref. \cite{nist} combined
with definition (\ref{def}). However, 
by using the parameters $\alpha$ and $\beta$ given
in (\ref{sph-2a-}), we find that Eqs. (\ref{using}) and
(\ref{convergencia-2}) are formally identical.
Therefore, $Y(y)=\sum\mathring {b}_n^1F_{n}(y)$ is also
solution of Eq. (\ref{V}).

\subsection{Convergence of the one-sided solutions}
{Before studying the convergence, notice that the function $\Psi(a,c;y)$ is satisfactory 
in the neighborhood of infinity but it may become infinity at $y=0$; 
in particular, it presents logarithmic terms 
at $y=0$ if $c=0,1,2$ \cite{nist}. Thus, we suppose that the solutions
$U_i^{\infty}$ are not valid when $y=0$. Furthermore, 
we will see that the the ratio test 
for convergence is inapplicable to solutions (\ref{che-primeiro-set})
when $z\to\infty$. For this reason, we consider the behaviour of each
solution when $z\to\infty$. In this manner, we will find that:
(i) the solutions $\mathring{U}_1$ and $\mathring{\textsf{U}}_1$
converge only for finite values of $z$, (ii) the solution 
$\mathring{U}_1^{\infty}$ converges
for $z\neq 0$; it converges also at $z=\infty$ only if 
$\text{Re}(i\eta-B_2/2)<-1$.} As to the other solutions, 
see the end of this subsection.

The convergence of an one-sided infinite series $\sum_{n=0}^{\infty}f_n(z)$
is obtained by using the limit
\begin{eqnarray}\label{ratio-1}
L_1(z)=\lim_{n\to\infty}\left|\frac{f_{n+1}(z)}{
f_{n}(z)}\right|.
\end{eqnarray}
By the D'Alembert ratio test, the series converges 
in the region where 
$L_1(z)<1$ and diverges where $L_1(z)>1$; if
$L_1=1$, the test is inconclusive. By the Raabe test \cite{watson,knopp}, if 
\begin{eqnarray}\label{Raabe-1}
L_1(z)=1+({A}/{n})
\end{eqnarray}
(where $A$ is constant), the series converges if 
$A<-1$ and diverges if $A>-1$; if  $A=-1$, the test is
inconclusive. 
First we use the Raabe test to get the convergence
of $\mathring{U}_1^{\infty}(z)$
and $\mathring{U}_1(z)$; second, we use the 
D'Alembert test to study the convergence of 
$\mathring{\textsf{U}}_1(z)$.

For $n\to\infty$ the recurrence relations for $\mathring{b}_n^{1}$
lead to
\begin{eqnarray}\label{min-1}
&&\begin{array}{l}
-2i\omega z_0\left[1+\frac{1}{n}\right]
\frac{\mathring{b}_{n+1}^{1}}{\mathring{b}_n^{1}}+
\left[n+1+2i\omega z_{0}-B_{2}-\frac{2B_1}{z_0}\right]
-\left[n+1+i\eta -\frac{2B_1}{z_0}-\frac{3B_2}{2}\right]
\frac{\mathring{b}_{n-1}^{1}}{\mathring{b}_n^{1}}=0,
\end{array}
\end{eqnarray}
whose solutions, when $n\to\infty$, satisfy
\letra
\begin{eqnarray}\label{minimal-bn}
\begin{array}{l}
  \frac{\mathring{b}_{n+1}^{1}}{\mathring{b}_n^{1}}\sim
1+\frac{1}{n}\left(i\eta-\frac{B_2}{2}\right)\quad
\Rightarrow\quad
\frac{\mathring{b}_{n-1}^{1}}{\mathring{b}_n^{1}}\sim
1-\frac{1}{n}\left(i\eta-\frac{B_2}{2}\right)\quad \text{or}
\end{array}
\end{eqnarray}
\begin{eqnarray}\label{minimal-bn'}
\begin{array}{l}
  \frac{\mathring{b}_{n+1}^{1}}{\mathring{b}_n^{1}}\sim\frac{n}{2i\omega z_0}\left[
1-\frac{1}{n}\left(B_2+\frac{2B_1}{z_0}\right)\right]\quad
\Rightarrow\quad
\frac{\mathring{b}_{n-1}^{1}}{\mathring{b}_n^{1}}\sim\frac{2i\omega z_0}{n}\left[
1+\frac{1}{n}\left(1+B_2+\frac{2B_1}{z_0}\right)\right]
\end{array}
\end{eqnarray}
On the other hand, from the last equation given in (\ref{using}) 
we find ($y=-2i\omega z$)
\antiletra
\begin{eqnarray} \label{relation-2}
\begin{array}{l}
\left[n+1+i\eta-\frac{B_1}{z_0}-\frac{B_2}{2}\right]
 \frac{\mathscr{F}_{n+1}(y)}{\mathscr{F}_{n}(y)}
-\left[n-1+2i\omega z-\frac{B_1}{z_0}\right]
+2i\omega z
\frac{\mathscr{F}_{n-1}(y)}{\mathscr{F}_{n}(y)}=0.
\end{array}
\end{eqnarray}
If $z$ is bounded (that is, if $2i\omega z/n\to 0$), 
then when $n\to\pm \infty$
($\ref{relation-2}$) is satisfied by two expressions for
${\mathscr{F}_{n+1}}/{\mathscr{F}_{n}}$ or
${\mathscr{F}_{n-1}}/{\mathscr{F}_{n}}$, namely,
\letra
\begin{eqnarray}\label{letra-1}
\begin{array}{lll}
\frac{\mathscr{F}_{n+1}}{\mathscr{F}_{n}}\sim 1+
\frac{1}{n}\left(\frac{B_2}{2}-2-i\eta\right)\quad& \Leftrightarrow
& \quad
\frac{\mathscr{F}_{n-1}}{\mathscr{F}_{n}}\sim
 1-\frac{1}{n}\left(\frac{B_2}{2}-2-i\eta\right),
\end{array}
\end{eqnarray}
\begin{eqnarray}\label{letra-2}
\begin{array}{lll} \frac{\mathscr{F}_{n+1}}{\mathscr{F}_{n}}\sim
 \frac{2i\omega z}{n}\left(1+\frac{B_1}{nz_0}\right)\qquad& \Leftrightarrow &\qquad
 \frac{\mathscr{F}_{n-1}}{\mathscr{F}_{n}}\sim
  \frac{n}{2i\omega z}\left[1-\frac{1}{n}\left(1+\frac{B_1}{z_0}\right)\right].
\end{array}
\end{eqnarray}
When $n\to\infty$, we have to choose the first possibility for both cases, that is, 
\begin{eqnarray}\label{possibilidade-1}
\begin{array}{l}
\displaystyle \lim_{n\to\infty}\frac{\mathscr{F}_{n+1}}{\mathscr{F}_{n}}= 1+
\frac{1}{n}\left(\frac{B_2}{2}-2-i\eta\right)\quad \text{for } 
\quad\mathscr{F}_{n}=\Psi \text{ and } \mathscr{F}_{n}=\tilde{\Phi}.
\end{array}
\end{eqnarray}
The ratio for $\tilde{\Phi}$ results from the fact that
$\lim_{c\to\infty}=\Phi(a,c;y)=1$
if $a$ and $y$ remain fixed and bounded \cite{erdelii}. 
For $\mathscr{F}_{n}=\Psi$ that choice is consistent 
with the fact that, if $|c|\to\infty$ while $a$
and $y$ remain fixed and bounded, then \cite{erdelii}
\antiletra
\begin{eqnarray}\label{criterio-1}
\begin{array}{l}
\Psi(a,c;y)\sim
c^{-a}\left[
(-1)^{-a}+\frac{\sqrt{2\pi}}{\Gamma(a)}
\ \left(\frac{c}{ey} \right) ^{c+a-\frac{3}{2}}y^{a-\frac{1}{2}}\ e^{y+a-\frac{3}{2}}\right]
\left[1+O\left(\frac{1}{|c|}\right)\right],
\end{array}\\
\left[c\to \infty; \ \ a\neq 0,-1,-2,\cdots;\ \ |\arg(\pm c)|<\pi\right]. 
\nonumber \ \
\end{eqnarray}
Thus, using (\ref{minimal-bn}) and (\ref{possibilidade-1}), 
we find
\begin{eqnarray}\label{raabe}
\begin{array}{l}
\displaystyle \lim_{n\to\infty}\frac{\mathring{b}_{n+1}^1\mathscr{F}_{n+1}}
{\mathring{b}_n^1\mathscr{F}_{n}}= 1-
\frac{2}{n}+O\left(\frac{1}{n^2}\right)\quad \text{for } 
\quad \stackrel{\infty}{U_1} \text{ and } U_1.
\end{array}
\end{eqnarray}
On the other hand, for $\mathring{\textsf{U}}_1$ 
we use the possibility corresponding to
(\ref{letra-2}), that is,  
%
%
%
\begin{eqnarray*}
\begin{array}{l}
\displaystyle\lim_{n\to+\infty} \frac{{F}_{n+1}(y)}{{F}_{n}(y)}=
 \frac{2i\omega z}{n}\left(1+\frac{B_1}{nz_0}\right),
\end{array}
\end{eqnarray*}
since this is the only one compatible with
\begin{eqnarray*}
\lim_{n\to\infty} {F}_{n}(y) &=&\begin{array}{l}
\frac{\Gamma\left[\frac{B_2}{2}-1-i\eta\right]}
{\Gamma\left[n-\frac{B_1}{z_0}\right]}
e^{-2i\omega z}
(2i\omega z)^{n-1-\frac{B_1}{z_0}}\end{array}\displaystyle
\lim_{n\to\infty}\begin{array}{l} {\Phi}\left(\frac{B_2}{2}-1-i\eta,
n-\frac{B_1}{z_0},-2i\omega z\right)
\end{array}\nonumber
\vspace{2mm}\\
&=&\begin{array}{l}
\frac{\Gamma[({B_2}/{2})-1-i\eta]}{\Gamma[n-({B_1}/{z_0})]}
e^{-2i\omega z}
(2i\omega z)^{n-1-\frac{B_1}{z_0}}\end{array}
\end{eqnarray*}
where we have used (\ref{kummer}). In this manner,
\begin{eqnarray}\label{dalambert}
\begin{array}{l}
\displaystyle \lim_{n\to\infty}\frac{\mathring{b}_{n+1}^1{F}_{n+1}}
{\mathring{b}_n^1{F}_{n}}=  \frac{2i\omega z}{n}+O\left(\frac{1}{n^2}\right)\quad \text{for } 
\quad \mathring{\textsf{U}}_1.
\end{array}
\end{eqnarray}

As the ratios (\ref{raabe}) and (\ref{dalambert}) are not valid at $z=\infty$, we
compute the asymptotic behaviour of the solutions.
So, from $\lim_{y\to \infty}\Psi(a,c;y)=y^{-a}$,  we find
\begin{eqnarray}
\lim_{z\to\infty}\mathring{U}_1^{\infty}(z)=e^{ i\omega z}\ z^{-i\eta-\frac{B_2}{2}}
\sum_{n=0}^{\infty}\mathring{b}_n^1, \qquad 
\lim_{n\to\infty}\left|\frac{\mathring{b}_{n+1}^{1}}{\mathring{b}_{n}^{1}}\right|
\stackrel{\text{(\ref{minimal-bn})}}{=}
1+\frac{1}{n}\text{Re}\left(i\eta-\frac{B_2}{2}\right).
\end{eqnarray}
Thus, by the Raabe test, the series in $\mathring{U}_1^ {\infty}$ converge 
at $z=\infty$ if $\text{Re}(i\eta-B_2/2)<-1$. 
%
%
%
To examine the asymptotic behaviour of  $\mathring{U}_1$
and $\mathring{\textsf{U}}_1$ 
we use \cite{nist} 
\begin{eqnarray*} \label{asymptotic-confluent}
&\begin{array}{l} \displaystyle
\lim_{y\to\infty}\tilde{\Phi}(a,c;y)\propto
\frac{\Gamma(c-a)}{\Gamma(a)}\,e^{y}y^{a-c}+ 
{e^{\pm i\pi a}y^{-a}},
\quad \left[a\neq 0,-1,\cdots; \; c-a\neq 0,-1,\cdots\right]
\end{array}
\end{eqnarray*}
where the upper sign holds for $-\pi/2<\arg{y}<3\pi/2$
and the lower sign, for $-3\pi/2<\arg{y}\leq-\pi/2$. Thence,  
we see that it is not possible to prove that $\mathring{U}_1$ and/or 
$\mathring{\textsf{U}}_1$ converge at $z=\infty$. We get, for example,
\begin{eqnarray*}
&\displaystyle\lim_{z\to\infty}\mathring{U}_1(z)\propto
e^{ i\omega z}\ z^{-i\eta-\frac{B_2}{2}}
\sum_{n=0}^{\infty}\mathring{b}_n^1
+e^{-i\omega z}\ z^{2+i\eta-\frac{B_1}{z_0}-
\frac{3}{2}B_2}
\sum_{n=0}^{\infty}
\begin{array}{l}\Gamma_n\;\mathring{b}_n^1\;(-2i\omega z) ^n,\end{array}
\\
&\begin{array}{l}
\Gamma_n=\Gamma\left(\frac{B_1}{z_0}+
\frac{B_2}{2}-n-i\eta\right),
\end{array}
\end{eqnarray*}
%
%
%
where the convergence of the second series becomes indefinite when $n\to\infty$
since 
\begin{eqnarray}\label{fi-infinito}
\lim_{n\to\infty}
\begin{array}{l}\frac{\Gamma_{n+1}\;\mathring{b}_{n+1}^1\,(-2i\omega z) ^{n+1}}
{\Gamma_n\;\mathring{b}_n^1\;(-2i\omega z) ^n}\;
\stackrel{\text{(\ref{minimal-bn})}}{=}\;
\frac{2i\omega z}{n}, \qquad (z\to\infty).\end{array}
\end{eqnarray}
%
%
%
%
%

 By using the transformations of the CHE as in (\ref{TRANS})
 we can extend the study of convergence for the other solutions.
 We find that:
\begin{itemize}
\itemsep-3pt
\item The one-sided infinite-series solutions $\mathring{U}_i$ and $\mathring{\textsf{U}}_i$
converge for finite values of $z$. 
\item
The one-sided infinite-series solutions $\mathring{U}_i^{\infty}$ 
in terms of irregular functions 
converge for finite values of $z$, excepting possibly the points $z= 0$ 
(if $i=1,2,3,4$)  
and $z=z_0$ (if $i=5,6,7,8$); the $\mathring{U}_i^{\infty}$ 
converge also at $z=\infty$ if 
\end{itemize}
\begin{equation}\label{regiao-1}
\begin{array}{ll}
\text{Re}\left[i\eta-\frac{B_2}{2}+1\right]< 0:\;\mathring{U}_1^{\infty},\,\mathring{U}_5^{\infty}; \qquad &
\text{Re}\left[i\eta-\frac{B_1}{z_0}-\frac{B_2}{2}\right]< 0:\;
\mathring{U}_2^{\infty},\,\mathring{U}_8^{\infty};\vspace{3mm}\\
\text{Re}\left[i\eta+\frac{B_2}{2}-1\right]< 0:\;\mathring{U}_3^{\infty},\,\mathring{U}_7^{\infty}; \qquad &
\text{Re}\left[i\eta+\frac{B_1}{z_0}+\frac{B_2}{2}\right]< 0:\; 
 \mathring{U}_4^{\infty},\,\mathring{U}_6^{\infty}.
\end{array}
\end{equation}
The above conditions also assure the convergence 
at $z=\infty$ of the two-sided series 
solutions ${U}_i^{\infty}$ discussed in the following.
%
%
%

%
%
%
%
%
%

%

 \section{Two-sided series solutions}
 In the present section, each set of the one-sided series 
 solutions ($\mathring{\mathbb{U}}_i$)  
 is transformed into two-sided series (${\mathbb{U}}_i$) by 
 replacing the summation index $n$ by $n+\nu_i$, where the 
 parameter $\nu_i$ must be chosen appropriately. 
 We write down only the first set of solutions 
 and establish the modifications 
in the domains of convergence which result
when we apply the ratio test for $n\to -\infty$.


 We use the convention
 %
 \begin{eqnarray}\label{notation-1}
 \mathbb{U}_i(z)=\left[U_{i}^{\infty}(z),{U}_{i}(z),{\textsf{U}}_{i}(z)\right]
 \end{eqnarray}
 and denote the respective series coefficients  by $b_n^{i}$; these satisfy
 recurrence relations having the form
  %
  \begin{eqnarray}\label{recursion-1}
  \alpha_{n}^{i}\  {b}_{n+1}^{i}+\beta_{n}^{i} \ {b}_n^{i}+
  \gamma_{n}^{i} \ {b}_{n-1}^{i}=0,
  \qquad [-\infty<n<\infty]
  \end{eqnarray}
 where $\alpha_{n}^{i}$, $\beta_{n}^{i}$ 
  and $\gamma_{n}^{i}$ depend on
  the parameters of the differential equation as well as on  
  the parameter $\nu_i$. This parameter $\nu_i$ 
  must be such that $\alpha_{n}^{i}\neq 0$
  and $\gamma_{n}^{i}\neq 0$, on the contrary the series
  truncates on the left ($\alpha_{n}^{i}= 0$) or on the 
  right-hand side ($\gamma_{n}^{i}= 0$) according to Appendix B. 
  From relations (\ref{recursion-1}) it results a transcendental (characteristic)
  equation given as a sum of two infinite continued fractions \cite{leaver}, namely,
  \begin{eqnarray}
  \label{recursion-2}
  \beta_{0}^{i}=\frac{\alpha_{-1}^{i}\gamma_{0}^{i}}{\beta_{-1}^{i}-}\ 
  \frac{\alpha_{-2}^{i}
  \gamma_{-1}^{i}}{\beta_{-2}^{i}-}\ 
  \frac{\alpha_{-3}^{i}\gamma_{-2}^{i}}
  {\beta_{-3}^{i}-}\cdots+\ 
  \frac{\alpha_{0}^{i}\gamma_{1}^{i}}{\beta_{1}^{i}-}
  \ \frac{\alpha_{1}^{i}\gamma_{2}^{i}}
  {\beta_{2}^{i}-}\ \frac{\alpha_{2}^{i}\gamma_{3}^{i}}{\beta_{3}^{i}-}\cdots.
  \end{eqnarray}
 If the differential equation has no free constant, the above equation 
 may be used to determine $\nu_i$; only for such values the solutions
 are valid. However, if the equation presents an arbitrary constant,
 we can ascribe convenient values for $\nu_i$; then, Eq. (\ref{recursion-2})
 permits to find values for that constant: in section IV, this fact is
 used to get bounded infinite-series
 eigenfunctions for the 
 QES potentials (\ref{potencial}).

 \subsection{The initial set of two-sided solutions}
 According to the the previous prescription, the first
 set of solutions is
 \begin{equation}\label{che-primeiro-set-nu}
 \begin{array}{l}
 {U}^{\infty}_1
 =\begin{array}{l}e^{i\omega z}z^{1+\frac{B_1}{z_0}}
 [z-z_0]^{1-B_2-\frac{B_1}{z_0}}\end{array}
 \displaystyle\sum_{n=-\infty}^{\infty}
 {b}_n^{1}
 \begin{array}{l}
 \Psi\left(2+i\eta-\frac{B_2}{2},2+\frac{B_1}{z_0}-n-\nu_1; - 2i\omega z
 \right)
 \end{array},
 \vspace{2mm}\\
 U_1
 =\begin{array}{l}e^{i\omega z}z^{1+\frac{B_1}{z_0}}
 [z-z_0]^{1-B_2-\frac{B_1}{z_0}}\end{array}
 \displaystyle\sum_{n=-\infty}^{\infty}
 {b}_n^{1}
 \begin{array}{l}
 \tilde{\Phi}\left(2+i\eta-\frac{B_2}{2},2+\frac{B_1}{z_0}-n-\nu_1; - 2i\omega z
 \right)
 \end{array},
 \vspace{2mm}\\
 {\textsf{U}}_1 = 
 \begin{array}{l}
 e^{i\omega z}
 [z-z_0]^{1-B_2-\frac{B_1}{z_0}}\end{array}
 \displaystyle\sum_{n=-\infty}^{\infty}
 {b}_{n}^{1}\left(2i\omega z\right)^{n+\nu_1}\vspace{2mm}\\
 \hspace{3.5cm}\times\begin{array}{l} \tilde{\Phi}\left(n+\nu_1+
 1+i\eta-\frac{B_1}{z_0}-\frac{B_2}{2}, %
 n+\nu_1-\frac{B_1}{z_0};-2i\omega z\right),\end{array}
 \end{array}
 \end{equation}
 with
 \begin{eqnarray}\label{alfa-nu}
 &&\begin{array}{l}
 {\alpha}_n^1=- 2i\omega z_0(n+\nu_1+1),\qquad
 \gamma_n^1=-\left[n+\nu_1+i\eta-\frac{B_1}{z_0}-\frac{B_2}{2}\right]
 \left[n+\nu_1+1-B_2-\frac{B_1}{z_0}\right],
 \end{array}\nonumber\\
 &&\begin{array}{l}
 {\beta}_n^1=(n+\nu_1)\left[n+\nu_1+1-B_2-\frac{2B_1}{z_0}+
 2i\omega z_0\right]+
 \left[i\omega z_0-1-\frac{B_1}{z_0}\right] \left[2-B_2-\frac{B_1}{z_0}
 \right]\end{array}\nonumber\\
 &&\begin{array}{l}
 \hspace{6mm}+2-B_2+B_3
 \end{array},
 \end{eqnarray}
 in the recurrence relations (\ref{recursion-1}). As in
 the one-sided series, ${2+i\eta-({B_2}/{2})
 \neq 0,-1,\cdots}$ for ${U}^{\infty}_1(z)$ and ${U}_1(z)$.
 In fact, the restrictions imposed on the parameters 
 of the hypergeometric functions of the the pairs 
 $(\mathring{U}^{\infty}_i,\mathring{U}_i)$ are valid also for 
 the two-sided solutions $({U}^{\infty}_i,{U}_i)$ .
 
 Solutions (\ref{che-primeiro-set-nu}) can be checked by 
 the same steps of section II.B, with  slight modifications. 
 For example, instead of Eq. (\ref{convergencia-1}) we find
 \begin{eqnarray*} 
 \displaystyle\sum_{n=-\infty}^{\infty}{b}_n^{1}
  {\alpha}_{n-1}^{1}\mathscr{F}_{n-1}(y)
 +\sum_{n=-\infty}^{\infty}{b}_n^{1}{\beta}_{n}^{1} \mathscr{F}_{n}(y)+
 \sum_{n=-\infty}^{\infty} {b}_n^{1}{\gamma}_{n+1}^{1}\mathscr{F}_{n+1}(y)=0
 \end{eqnarray*}
 where now
 \begin{eqnarray*}
 \begin{array}{l}
\mathscr{F}_n(y)=\Psi\left(2+i\eta-\frac{B_2}{2},2+
\frac{B_1}{z_0}-n-\nu_1;y\right) \text{ or } \end{array}
\begin{array}{l}
 \mathscr{F}_n(y)=\tilde{\Phi}
 \left(2+i\eta-\frac{B_2}{2},2+\frac{B_1}{z_0}-n-\nu_1;y\right)
 \end{array}\end{eqnarray*}
For two-sided series ($-\infty<n<\infty$), the above equation  
can be written as 
 \begin{eqnarray*}
 \displaystyle\sum_{n=-\infty}^{\infty}
 \left[{\alpha}_{n}^{1}\; {b}_{n+1}^{1}+
 {\beta}_{n}^{1} \;{b}_n^{1}+
 {\gamma}_{n}^{1} \;{b}_{n-1}^{1}\right]
 \mathscr{F}_{n}(y)=0,
 \end{eqnarray*}
 which is satisfied by the recurrence relations (\ref{recursion-1}) with $i=1$.

 \subsection{Convergence of the two-sided series}
 
 For the series $\sum_{n=-\infty}^{\infty}f_n(z)$, in addition
 to the ratio (\ref{ratio-1}), we need to compute the ratio
 \begin{eqnarray}\label{ratio-2}
 L_2(z)=\lim_{n\to-\infty}\Big|{f_{n-1}(z)}/{
 f_{n}(z)}\Big|.
 \end{eqnarray}
 By the Raabe test, if 
 \begin{eqnarray}\label{Raabe-1}
 L_2(z)=1+\big({B}/{|n|}\big)
 \end{eqnarray}
 (where $B$ is a constant) the series converges when 
 $B<-1$ and diverges when $B>-1$; for $B=-1$ the test is
 inconclusive. Hence, the convergence of two-sided series is given by 
 the intersection of the region obtained when $n\to\infty$ with the region
 obtained when $n\to-\infty$.
 
 When $n\to\infty$ the ratio test does not alter the convergence, that is,
 the regions stated at the end of section II.D hold for two-sided 
 expansions as well. On the other hand, when $n\to-\infty$, the 
 ${U}_{i}(z)$ converge for any finite value of $z$ as $\mathring{U}_{i}(z)$ do, 
 but the domains of convergence of ${U}_{i}^{\infty}(z)$ and  ${\textsf{U}}_{i}(z)$
 are restricted. Then, if the parameters of the confluent hypergeometric 
 functions satisfy the same conditions as in the case of one-sided
 solutions, the actual regions of convergence are given by:
\begin{itemize}
\itemsep-3pt
\item The two-sided solutions ${U}_i$ 
converge for any finite value of $z$; the ${\textsf{U}}_{i}$ converge
only for the finite values of $z$ which satisfy the restrictions  
given in (\ref{convergencia1-nu1}) and (\ref{convergencia1-nu2}). 
\item
The two-sided solutions ${U}_i^{\infty}$ 
converge at $z=\infty$ if the conditions (\ref{regiao-1}) are fulfilled; 
these ${U}_{i}^{\infty}$ converge
only for finite values of $z$ which satisfy the restrictions 
specified in (\ref{convergencia1-nu1}) and (\ref{convergencia1-nu2}).
\end{itemize}
  \begin{equation}\label{convergencia1-nu1}
  |z|\geq|z_0|  \text{ if }
  \begin{cases}\text{Re}\left[ B_2+\frac{B_1}{z_0}\right] {>}1 \text{ for }
  ({U}_{1}^{\infty},{\textsf{U}}_1) \text{ and } ({U}_{2}^{\infty},{\textsf{U}}_{2}),\vspace{2mm} \\
  \text{Re}\left[  B_2+\frac{B_1}{z_0}\right] {<}1 \text{ for } 
   ({U}_{3}^{\infty},{\textsf{U}}_3)
   \text{ and } ({U}_{4}^{\infty},{\textsf{U}}_4),
  \end{cases} 
  \end{equation}
  \begin{equation}\label{convergencia1-nu2}
  |z-z_0|\geq|z_0| \text{ if }
  \begin{cases}\text{Re}\left[ \frac{B_1}{z_0}\right] {<}-1 \text{ for } 
  ({U}_{5}^{\infty},{\textsf{U}}_5) \text{ and } ({U}_{8}^{\infty},{\textsf{U}}_8),
  \vspace{2mm} \\
  \text{Re}\left[ \frac{B_1}{z_0}\right] {>}-1 \text{ for } 
  ({U}_{6}^{\infty},{\textsf{U}}_6) \text{ and } ({U}_{7}^{\infty},
  {\textsf{U}}_7), 
  \end{cases}
  \end{equation}
  where the constraints on $B_2+(B_1/z_0)$ or $B_1/z_0$ 
  are necessary only to assure the convergence at $|z|=|z_0|$ 
 or $|z-z_0|=|z_0|$, respectively, that is: without these constraints, 
 the solutions indicated in 
 (\ref{convergencia1-nu1}) converges for $|z|>|z_0|$, while 
the ones indicated in (\ref{convergencia1-nu1}) converges for $|z-z_0|>|z_0|$.

{Notice that, if for a given problem a solution $U_i^{\infty}$
converges at $z=\infty$, then this $U_i^{\infty}$ and $U_i$ cover the entire
complex plane $z$. This fact does not occur with the two-sided expansions
in series of Coulomb wave functions \cite{lea-2013}.}

  Let us find the restrictions (\ref{convergencia1-nu1}) and (\ref{convergencia1-nu2}).
  It is sufficient to consider explicitly only
  the firts set; the results for the other sets
  follow by the transformations (\ref{transformacao1}). 
  In the first place, for $n\to\pm\infty$ 
  insted of (\ref{min-1}), we have
\begin{equation*}
\begin{array}{l}
2i\omega z_0\left[1+\frac{1}{n}\right]
\frac{{b}_{n+1}^{1}}{{b}_n^{1}}-
\left[n+\nu_1+1+2i\omega z_{0}-B_{2}-\frac{2B_1}{z_0}\right]
+\left[n+\nu_1+1+i\eta -\frac{2B_1}{z_0}-\frac{3B_2}{2}\right]
\frac{{b}_{n-1}^{1}}{{b}_n^{1}}=0,
\end{array}
\end{equation*}
whose minimal solution behaves as
 \begin{eqnarray}\label{behaves}
 \begin{array}{l}
 \frac{{b}_{n-1}^{1}}{{b}_n^{1}}\sim\frac{2i\omega z_0}{n}\left[
 1+\frac{1}{n}\left(1+B_2+\frac{2B_1}{z_0}-\nu_1\right)\right] 
 \quad
 \Rightarrow\quad
 \frac{{b}_{n+1}^{1}}{{b}_n^{1}}\sim\frac{n}{2i\omega z_0}\left[
 1-\frac{1}{n}\left(B_2+\frac{2B_1}{z_0}-\nu_1\right)\right],
 \end{array}
 \end{eqnarray}
On the other hand, relation (\ref{relation-2}) is replaced by  ($y=-2i\omega z$)
\begin{eqnarray} \label{relation-2-nu}
\begin{array}{l}
\left[n+\nu_1+1+i\eta-\frac{B_1}{z_0}-\frac{B_2}{2}\right]
 \frac{\mathscr{F}_{n+1}(y)}{\mathscr{F}_{n}(y)}
-\left[n+\nu_1-1+2i\omega z-\frac{B_1}{z_0}\right]
+2i\omega z
\frac{\mathscr{F}_{n-1}(y)}{\mathscr{F}_{n}(y)}=0.
\end{array}
\end{eqnarray}
where, now, 
\begin{eqnarray*}\begin{array}{l}
\mathscr{F}_n(y)=\Psi\left(2+i\eta-\frac{B_2}{2},2+\frac{B_1}{z_0}-n-\nu_1;y\right) \text{ or } 
\mathscr{F}_n(y)=\tilde{\Phi}\left(2+i\eta-\frac{B_2}{2},2+\frac{B_1}{z_0}-n-\nu_1;y\right).
\end{array}
\end{eqnarray*}
If $z$ is bounded (that is, if $2i\omega z/n\to 0$), 
then Eq. (\ref{relation-2-nu}) when $n\to\pm\infty$ gives
%
\begin{eqnarray*}\label{letra-1-nu}
\begin{array}{lll}
\frac{\mathscr{F}_{n+1}}{\mathscr{F}_{n}}\sim 1+
\frac{1}{n}\left(\frac{B_2}{2}-2-i\eta\right)\quad& \Leftrightarrow
& \quad
\frac{\mathscr{F}_{n-1}}{\mathscr{F}_{n}}\sim
 1-\frac{1}{n}\left(\frac{B_2}{2}-2-i\eta\right),
\end{array}
\end{eqnarray*}
\begin{eqnarray*}\label{letra-2-nu}
\begin{array}{lll} \frac{\mathscr{F}_{n+1}}{\mathscr{F}_{n}}\sim
 \frac{2i\omega z}{n}\left[1+\frac{1}{n}\left(\frac{B_1}{z_0}-\nu_1\right)\right]
 \qquad& \Leftrightarrow &\qquad
 \frac{\mathscr{F}_{n-1}}{\mathscr{F}_{n}}\sim
  \frac{n}{2i\omega z}\left[1-\frac{1}{n}\left(1-\nu_1+\frac{B_1}{z_0}\right)\right].
\end{array}
\end{eqnarray*}
 For $n\to -\infty$ we have to select 
 \begin{equation}
 \begin{array}{l}
  \frac{\mathscr{F}_{n-1}}{\mathscr{F}_{n}}\sim
  \frac{n}{2i\omega z}\left[1+\frac{1}{n}\left(\nu_1-1-\frac{B_1}{z_0}\right)\right]\text{ if }
  \mathscr{F}_n=\Psi,\quad
  \frac{\mathscr{F}_{n-1}}{\mathscr{F}_{n}}\sim
 1-\frac{1}{n}\left(\frac{B_2}{2}-2-i\eta\right)\text{ if }
  \mathscr{F}_n=\tilde{\Phi}
 \end{array}
 \end{equation}
 Thence, using also the ratio (\ref{behaves}), we find 
  \begin{eqnarray*}\begin{array}{l}\displaystyle
  \lim_{n\to -\infty}\frac{{b}_{n-1}^{1}\mathscr{F}_{n-1}}{{b}_n^{1}\mathscr{F}_{n}}=\frac{z_0}{z}
 \left[1-\frac{1}{|n|}\left(B_2+\frac{B_1}{z_0}\right)\right]+O\left(\frac{1}{n^2}\right),
 \qquad [ \mathscr{F}_n=\Psi \mapsto U_1^{\infty}]
 \end{array}
 \end{eqnarray*}
 \begin{eqnarray*}\begin{array}{l}\displaystyle
  \lim_{n\to -\infty}\frac{{b}_{n-1}^{1}\mathscr{F}_{n-1}}{{b}_n^{1}\mathscr{F}_{n}}=
 \frac{2i\omega z_0}{n}+O\left(\frac{1}{n^2}\right), 
 \qquad [ \mathscr{F}_n=\tilde{\Phi}\mapsto U_1].
 \end{array}
 \end{eqnarray*}
The first relation leads to the restriction (\ref{convergencia1-nu1}) 
for $ U_1^{\infty}$; 
the second relation does not restrict the convergence of $U_1$. 
To show that
the convergence of ${\textsf{U}}_i$ is restricted as in  
(\ref{convergencia1-nu1}), we define $F_n(y)$ as
 \begin{eqnarray*}
\begin{array}{l}
 F_n(y)=(2i\omega z)^{n+\nu_1}\tilde{\Phi}\left(n+\nu_1+
 1+i\eta-\frac{B_1}{z_0}-\frac{B_2}{2},
 n+\nu_1-\frac{B_1}{z_0};-2i\omega z\right). 
 \end{array}
 \end{eqnarray*}
 This $F_n$ satisfies (\ref{relation-2-nu}) which, for $n\to -\infty$, gives
 \begin{equation*}
 \begin{array}{l}
  \frac{{F}_{n-1}}{{F}_{n}}\sim
  \frac{n}{2i\omega z}\left[1+\frac{1}{n}\left(\nu_1-1-\frac{B_1}{z_0}\right)\right].
 \end{array}
 \end{equation*}
 Together with the ratio (\ref{behaves}) the previous limit
 produces the condition given in (\ref{convergencia1-nu1}).
{ On the other side, by using  $\lim_{y\to \infty}\Psi(a,c;y)=y^{-a}$,  we find
\begin{eqnarray*}
\lim_{z\to\infty}{U}_1^{\infty}(z)=e^{ i\omega z}\ z^{-i\eta-\frac{B_2}{2}}
\sum_{n=-\infty}^{\infty}{b}_n^1, \qquad 
\lim_{n\to-\infty}\left|\frac{{b}_{n-1}^{1}}{{b}_{n}^{1}}\right|
\stackrel{\text{(\ref{behaves})}}{=}
\left|\frac{2i\omega z_0}{n}\right|\to 0.
\end{eqnarray*}
 Therefore, $U_1^{\infty}$ converges at $z=\infty$ if the first condition
 given in (\ref{regiao-1}) is satisfied. }


\section{Schr\"{o}dinger equation for hyperbolic potentials}
For the hyperbolic potential (\ref{potencial}),
the Schr\"{o}dinger equation (\ref{schr}) becomes
  \begin{eqnarray*}
  \label{schr-1a}
  \frac{d^2\psi}{du^2}+\Big\{{\cal E}-
  4\beta^{2}{\sinh^{4}u}- 4\beta\big[\beta-
    2(\gamma+\delta+\ell)\big]{\sinh^{2}u}\nonumber\\
\begin{array}{l}
-4\left[ \delta-\frac{1}{4}\right] \left[\delta-\frac{3}{4}\right]\frac{1}{\sinh^{2}u}
    +4\left[ \gamma-\frac{1}{4}\right] \left[\gamma-\frac{3}{4}\right]\frac{1}
      {\cosh^{2}u}
  \Big\}\end{array}\psi=0. 
  \end{eqnarray*}
The substitutions
\begin{eqnarray} \label{Ush}
z=\cosh^2u,\qquad \psi(u)={\psi}[u(z)]=
z^{\gamma-\frac{1}{4}}(z-1)^{\delta-\frac{1}{4}}U(z),\qquad[z\geq 1]
\end{eqnarray}
transform the preceding equation 
into a confluent Heun equation with 
\begin{eqnarray}\label{parametros-ush}
&\begin{array}{l}
z_0=1,\qquad B_{1}=-2\gamma, \qquad  B_{2}=2\gamma+2\delta, \qquad 
B_{3}=\frac{{\cal E}}{4}+\left(\gamma+\delta-\frac{1}{2}\right)^{2},\end{array} 
\vspace{2mm}\nonumber\\
&i\omega=\pm\beta,\qquad i\eta= \pm(\ell+\delta+\gamma),
\end{eqnarray}
where the upper or the lower sign for ($\eta,\omega)$ must 
be taken throughout.

%
%
%
%

\subsection{The Ushveridze potential} 
Now we exclude the four cases
given in Eq. (\ref{4casos}) and suppose that $\ell$ is a non negative
integer. Due to the factor $\exp(i\omega z)$, 
in order to obtain bounded solutions when
$z\to\infty$ we take
\[i\omega=-\beta,\qquad i\eta=-\ell-\gamma-\delta, \qquad[\ell=0,1,2,\cdots].\]
in Eq. (\ref{parametros-ush}). 

When $\delta\geq1/4$, the above family 
of potentials is quasiexactly
solvable because it admits
bounded wavefunctions given by finite series which   
allow to determine only a finite
number of energy levels for any real value of the
parameter $\gamma$. In addition, 
for $1/4\leq\delta<1/2$ and $1/2<\delta\leq 3/4$
we have found two-sided infinite series of Coulomb
wave functions which are
convergent and bounded for any $z\geq 1$: this 
gives the possibility of determining
the remaining part of
the energy spectra as solutions of a
characteristic equation. {For $\delta=1/2$, the Raabe test is inconclusive.
We will show that these 
results are more general than the ones obtained from the new 
two-sided solutions
for the CHE.}
%

We will see that it is advantageous to use the Baber-Hass\'e expansions
to get finite-series solutions. 
%
%
%
%
Then, by inserting the solution $U_1^{\text{baber}}$
given in (\ref{barber-1}) into Eq. (\ref{Ush}), 
we find 
\letra
\begin{eqnarray} \label{baber}
\psi_1^{\text{baber}}[u(z)]=
e^{-\beta z}z^{\gamma-\frac{1}{4}}(z-1)^{\delta-\frac{1}{4}}
\sum_{n=0}^{\ell}a_n^1(z-1)^n,\qquad \delta\geq\frac{1}{4}
\end{eqnarray}
where the coefficients satisfy the relations ($a_{-1}^1=0$)
\begin{eqnarray}\label{baber-I}
\begin{array}{l}
A_n^1\;a_{n+1}^1+ 
B_n^1\;a_n^1+
G_n^1\;a_{n-1}^1=0,\qquad 0\leq n\leq \ell \quad \text{with}\end{array}
\end{eqnarray}
\begin{eqnarray*}
&\begin{array}{l}
A_n^1=(n+1)(n+2\delta),\quad 
B_n^1=n(n+2\gamma+2\delta-1-2\beta)+\frac{{\cal E}}{4}+
\left(\gamma+\delta-\frac{1}{2}\right)^2 -2\beta\delta,
\end{array}\nonumber\vspace{2mm}\\
&G_n^1=-2\beta(n-\ell-1).
\end{eqnarray*}
According to Appendix A, the series in (\ref{baber}) ends at 
$n=\ell$ because $G_n^1=0$ when $n=\ell+1$.  
Since $\beta>0$, the previous eigenfunctions are bounded  
for all $z\geq 1$ (including $z=1$) provided that 
$\delta\geq 1/4$. In fact,  the condition
$A_n^1\;G_{n+1}^1> 0$
($0\leq n\leq \ell-1$) assures
that $\psi_1^{\text{baber}}$ represents $\ell+1$ solutions, 
each one with a different real energy 
determined from the vanishment of the determinant
of the tridiagonal matrix (\ref{matriz-tridiagonal}) 
corresponding to (\ref{baber-I}) \cite{arscott}.  

From (\ref{hiper-baber}) we see that there is a
finite-series solution $\mathring{\psi}_4$ 
associated with the the expansion $\mathring{\textsf{U}}_4$ in series 
of confluent hypergeometric functions:
\begin{eqnarray*} 
\mathring{\psi}_4[u(z)]=
e^{-\beta z}z^{\gamma-\frac{1}{4}}(z-1)^{\delta-\frac{1}{4}}
\sum_{n=0}^{\ell}\mathring{b}_n^4(-2\beta z)^n\tilde{\Phi}
\left(n-\ell,n+2\gamma;2\beta z\right), 
\end{eqnarray*}
where the $\mathring{b}_n^4$ satisfy (\ref{r1a}) with
\begin{eqnarray*}
\mathring{\alpha}_n^4=2\beta(n+1),\qquad 
\mathring{\beta}_n^4=B_n^1,
\qquad\mathring{\gamma}_n^4=-\left[n-1+2\delta\right]
\left[n-\ell-1\right],\qquad 
\end{eqnarray*}
Contrary to (\ref{baber-I}), these wavefunctions are valid only
if $2\gamma\neq 0,-1,-2,\cdots$. In this sense, 
the Baber-Hass\'e solutions
 are more general when we are concerned with 
 finite-series solutions, but they do not afford infinite series 
convergent at $z=\infty$. 

To get infinite series we use the two-sided 
solutions. 
We impose the conditions
(\ref{regiao-1}) and (\ref{convergencia1-nu1}), and require that $a$ 
is not zero or a negative integer in $\Psi(a,c;y)$. Thus,
for $U_2^{\infty}$ Eq. (\ref{Ush}) becomes
\antiletra\letra
\begin{eqnarray} 
\begin{array}{l}{\psi}_2[u(z)]=
e^{-\beta z}z^{\gamma-\frac{1}{4}}(z-1)^{\frac{3}{4}-\delta}\displaystyle
\sum_{n=-\infty}^{\infty}{b}_n^2\Psi
\left(1-\ell-2\delta,2\gamma-n-\nu_2;2\beta z\right), 
\end{array}\\
\begin{array}{l} [{1}/{2}<\delta\leq{3}/{4}]\end{array}\nonumber
\end{eqnarray}
where the $b_n^2$ satisfy (\ref{recursion-1}) with
\begin{eqnarray}
&\begin{array}{l}
{\alpha}_n^2=2\beta(n+\nu_2+1),\qquad 
\qquad{\gamma}_n^2=-\left[n+\nu_2+1-2\gamma-2\delta\right]
\left[n+\nu_2+1-\ell-2\delta\right],\end{array}\nonumber\\
&\begin{array}{l}
{\beta}_n^2=[n+\nu_2][n+\nu_2+3-2\gamma-2\delta-2\beta]+\frac{{\cal E}}{4}+
\left[\gamma+\delta-\frac{1}{2}\right]^2 -
2\beta[1-\delta]+2-2\gamma-2\delta.\end{array}\nonumber\\
\end{eqnarray}
 According to (\ref{convergencia1-nu1}), the series converges 
  at $z=1$ if $\delta>1/2$; in addition, the solution is bounded at $z=1$ if $\delta\leq3/4$: 
	thence the series converges also 
 at $z=\infty$ because in such cases $\ell+2\delta>0$ 
  as required by (\ref{regiao-1}). Furthermore, to have 
  two-sided infinite series,  
the parameter $\nu_2$ cannot allows that 
${\alpha}_n^2$ and ${\gamma}_n^2$ vanish, that is,
\begin{eqnarray}
\nu_2,\qquad\nu_2-2\gamma-2\delta,\qquad \nu_2-2\delta\qquad \text{are not integers}.\end{eqnarray}
On the other side, using $U_4^{\infty}$ we find
\antiletra\letra
\begin{eqnarray} \label{psi-4}
\begin{array}{l}{\psi}_4[u(z)]=
e^{-\beta z}z^{\frac{3}{4}-\gamma}(z-1)^{\delta-\frac{1}{4}}\displaystyle
\sum_{n=-\infty}^{\infty}{b}_n^4\Psi
\left(1-\ell-2\gamma,2\gamma-n-\nu_4;2\beta z\right), 
\end{array}\\
\begin{array}{l}
[1/4\leq\delta<{1}/{2}, \; \ell+2\gamma>0,\; \ell+2\gamma\neq 1,2,3,\cdots]
\nonumber
\end{array}
\end{eqnarray}
where the $b_n^4$ satisfy (\ref{recursion-1}) with
\begin{eqnarray}
&\begin{array}{l}
{\alpha}_n^4=2\beta(n+\nu_4+1),\qquad 
\qquad{\gamma}_n^4=-\left[n+\nu_4-1+2\delta\right]
\left[n+\nu_4-1-\ell\right],\end{array}\nonumber\nonumber\\
&\begin{array}{l}
{\beta}_n^4=[n+\nu_4][n+\nu_4-1+2\gamma+2\delta-2\beta]+\frac{{\cal E}}{4}+
\left[\gamma+\delta-\frac{1}{2}\right]^2 -2\beta\delta.\end{array}
\end{eqnarray}
The condition $\delta<1/2$ assures that the series converges 
at $z=1$, while $1/4\leq\delta$ assure that the solution is bounded at $z=1$; 
the condition $\ell+2\gamma>0$ assures convergence at $z=\infty$,
while $\ell+2\gamma\neq1,\,2,\,3,\cdots$ guarantees that
$\Psi (1-\ell-2\gamma,c;2\beta z)$ is not a polynomial 
of fixed degree. To have two-sided infinite series, 
 $\nu_4$ cannot allows that ${\alpha}_n^4$ and ${\gamma}_n^4$ vanish,
 a condition that is fulfilled if
 \begin{eqnarray}
 \nu_4 \text{ and } \nu_4+2\delta \text{ are not integers}.
 \end{eqnarray}
{The restrictions on the parameter $\gamma$ are sufficient to
see that the solutions (\ref{psi-4}) are less general than the expansions 
in series of Coulomb wave functions. }

\subsection{Whittaker-Hill equation for a hyperbolic Razavy potential}

Let us consider only the case $\gamma=\delta=1/4$ in (\ref{potencial}). 
We get
\antiletra
\begin{eqnarray*} 
\mathcal{V}(u)=4\beta^{2}{\sinh^{4}u}+ 4\beta\big[\beta-
1-2\ell \big]{\sinh^{2}u}, 
\qquad \ell=0,\ {1}/{2},\ 1,\ {3}/{2}, \cdots
\end{eqnarray*} 
where $u\in(-\infty,\infty)$. The substitutions
\letra
 \begin{eqnarray}
 z=\cosh^2u,\qquad \psi(u)={\psi}[u(z)]=U(z),
 \qquad[z\geq 1]
 \end{eqnarray}
 transform the Schr\"odinger equation for the preceding potential into
 the CHE (\ref{gswe}) with 
 \begin{eqnarray}\label{whe-parametros}
\begin{array}{l}
 z_0=1,\quad B_{1}=-\frac{1}{2}, \quad B_{2}=1, \quad 
 B_{3}=\frac{{\cal E}}{4},\qquad
 i\omega=\pm\beta,\quad i\eta= \pm\left(\ell+\frac{1}{2}\right),
 \end{array}
 \end{eqnarray}
that is, into the Whittaker-Hill equation (\ref{whe}).
To obtain bounded solutions for $z=\cosh^2u\geq 1$ we will choose
\[i\omega=-\beta,\qquad i\eta=-\ell-({1}/{2}).\]

By using expansions in series of Coulomb wave functions we have 
seen that, for $\ell$ integer or half-integer, this problem admits: 
(i) even and odd finite-series solutions (with respect to $u\mapsto -u$) which are 
 convergent and bounded for all $z\geq1$, 
(ii) even infinite-series solutions which are convergent and bounded for
 all $z\geq1$. {To find odd infinite-series solutions 
it is necessary to use two solutions valid for different intervals of $z$.
Below we find similar results by using the solutions for the CHE given in 
the present study. }

{In fact, there are several possibilities
to form finite-series solutions for this case. One can use, for example, the 
one-sided series solutions $\mathring{\textsf{U}}_i$ with $i=1,\cdots,4$ as
explained at the end of section II.B; notice that  
we cannot use the $\mathring{\textsf{U}}_i$ if $i=5,\cdots,8$, because these do not satisfy 
the conditions $\alpha_n^i\gamma_{n+1}^i>0$
which assure real energies \cite{arscott}. The final result is:
$\mathring{\textsf{U}}_1$ and $\mathring{\textsf{U}}_3$ yield, respectively, 
odd and even finite-series solutions when $\ell$ is half-integer;
if $\ell$ is a non-negative integer we can use
$\mathring{\textsf{U}}_2$ (odd) and $\mathring{\textsf{U}}_4$ (even).}

{To find even infinite-series solutions bounded
for any $z=\cosh^2u\geq 1$ we use the two-sided expansions 
${U}_3^{\infty}$ and ${U}_4^{\infty}$
for $\ell$ half-integer and integer, respectively: these satisfy 
the conditions (\ref{regiao-1}) and (\ref{convergencia1-nu1}) at
the same time and, consequently, converge at $z=\infty$ and $z=1$.}
 Thus 
 \antiletra\letra
 \begin{eqnarray}
 \begin{array}{l}
 {U}^{\infty}_3[z(u)]
 =e^{-\beta \cosh^2u}\end{array}
 \displaystyle\sum_{n=-\infty}^{\infty}
 {b}_n^{3}\begin{array}{l}
 \Psi\left(-\ell,\frac{1}{2}-n-\nu_3; 2\beta\cosh^2u
 \right),\quad
 \left[\ell=\frac{1}{2},\;\frac{3}{2},\;\frac{5}{2},\;\cdots\right]\end{array}
 \end{eqnarray}
 with
 \begin{eqnarray}
 &\begin{array}{l}
 {\alpha}_n^3=2\beta(n+\nu_3+1),\qquad
 {\beta}_n^3=(n+\nu_3)\left[n+\nu_3+1-2\beta\right]
 +\frac{1}{2}\left(\frac{1}{2}-\beta\right) +\frac{{\cal E}}{4},
 \end{array}\nonumber\\
 &\begin{array}{l}
 \gamma_n^3=-\left[n+\nu_3-\ell-\frac{1}{2}\right]
 \left[n+\nu_3-\frac{1}{2}\right],
 \end{array}
 \end{eqnarray}
 {in the recurrence relations (\ref{recursion-1}).
 The summation runs from $-\infty$ to $+\infty$
 if we choose $\nu_3$ in such a way that ${\alpha}_n$ 
 and ${\gamma}_n$ do not vanish, for example, $\nu_3\in(0,1/2)$.
 On the other side, if $\ell$ is integer, we find the infinite series}
 \antiletra\letra
 \begin{eqnarray}
 \begin{array}{l}
 {U}^{\infty}_4[z]=\cosh{u}\,
e^{-\beta\cosh^2u} \end{array}
 \displaystyle\sum_{n=-\infty}^{\infty}\begin{array}{l}
 {b}_n^{4}\Psi\left(\frac{1}{2}-\ell,\frac{3}{2}-n-\nu_4; 2\beta\cosh^2u\right),
 \end{array}\quad
 \left[\begin{array}{l}\ell=0,\;1,\;2,\;\cdots\end{array}\right]\;
 \end{eqnarray}
 with
 \begin{eqnarray}
 &\begin{array}{l}
 {\alpha}_n^4=2\beta(n+\nu_4+1),\qquad
 {\beta}_n^4=(n+\nu_4)\left[n+\nu_4-
 2\beta\right]-\frac{1}{2}\beta +\frac{{\cal E}}{4},
 \end{array}\nonumber\\
 &\begin{array}{l}
 \gamma_n^4=-\left[n+\nu_4-\ell-1\right]
 \left[n+\nu_4-\frac{1}{2}\right]
 \end{array}
 \end{eqnarray}
{in the recurrence relations (\ref{recursion-1}). The parameter
 $\nu_4$ can be chosen as above, $\nu_4\in(0,1/2)$.}

{Odd infinite-series solutions 
are obtained if we use two-sided solutions 
which are valid for different ranges of $z$ but
have the same characteristic equation, namely:
the sets $\mathbb{U}_1$ and $\mathbb{U}_2$
for $\ell$ half-integer and integer, respectively. Thus, 
as an example we suppose that $\ell$ is half-integer. Then, 
we have the odd set }
\antiletra\letra
\begin{eqnarray}
\begin{array}{l}
\begin{array}{l}
{U}^{\infty}_1[z(u)]
=\sinh(2u)\:
e^{-\beta \cosh^2u}\end{array}
\displaystyle\sum_{n=-\infty}^{\infty}
{b}_n^{1}\begin{array}{l}
\Psi\left(1-\ell,\frac{3}{2}-n-\nu_1; 2\beta\cosh^2u
\right),\end{array}\vspace{2mm} \\
\begin{array}{l}
{U}_1[z(u)]
=\sinh(2u)\:
e^{-\beta \cosh^2u}\end{array}
\displaystyle\sum_{n=-\infty}^{\infty}
{b}_n^{1}\begin{array}{l}
\tilde{\Phi}\left(1-\ell,\frac{3}{2}-n-\nu_1; 2\beta\cosh^2u
\right), 
%
\end{array}\vspace{2mm}\\
\begin{array}{l}
{\textsf{U}}_1[z(u)] = \sinh{u}\,e^{-\beta \cosh^2u}\end{array}
\displaystyle\sum_{n=-\infty}^{\infty}{b}_n^{1}\left[-2\beta\cosh^2u\right]^{n+\nu_i}\times\\
\hspace{5.5cm}
\begin{array}{l}
\tilde{\Phi}\left(n+\nu_1-\ell+\frac{1}{2},n+\nu_1+\frac{1}{2}; 2\beta\cosh^2u
\right),
\end{array} 
\end{array}
\end{eqnarray}
with
\begin{eqnarray}
&\begin{array}{l}
{\alpha}_n^1=2\beta(n+\nu_1+1),\qquad
{\beta}_n^1=(n+\nu_1)\left[n+\nu_1+1-2\beta\right]
-\frac{3}{2}\beta +\frac{{\cal E}+1}{4},
\end{array}\nonumber\\
&\begin{array}{l}
\gamma_n^1=-\left[n+\nu_1-\ell-\frac{1}{2}\right]
\left[n+\nu_1+\frac{1}{2}\right],
\end{array}
\end{eqnarray}
{in the recurrence relations (\ref{recursion-1}).
The summation runs from $-\infty$ to $+\infty$
if we choose $\nu_1$ in such a way that ${\alpha}_n$ 
and ${\gamma}_n$ do not vanish, for example, $\nu_1\in(0,1/2)$.
The solution $U_1^{\infty}$ converges at $z=\cosh^2{u}=\infty$ because 
it satisfies the condition (\ref{regiao-1}) ($\ell>0$); it does not 
converge at $z=1$ because the condition given (\ref{convergencia1-nu1}) 
is not fulfilled. On the other hand, the solution ${U}_1$
is valid for any finite value of $z$. Therefore, these two solutions 
cover all of the admissible values for $z$; furthermore, by means of 
relation (\ref{continuacao}), 
$\textsf{U}_1$ can be written as a linear combination of $U_1^{\infty}$ and
${U}_1$ in the region where $z\neq1$ and $z\neq\infty$.}

%
%
%
%
%
%

\section{Schr\"{o}dinger equation for trigonometric potentials}
For the trigonometric potential (\ref{ush-trigonometrico}),
the Schr\"{o}dinger equation (\ref{schr}) becomes
\antiletra
  \begin{eqnarray*}
  \label{schr-2a}
&&\frac{d^2\psi}{du^2}+\Big\{{\cal E}+
  4\beta^{2}{\sin^{4}u}- 4\beta\big[\beta+
    2(\gamma+\delta+\ell) \big]{\sin^{2}u}\nonumber\\
&&\begin{array}{l}
-4\left[ \delta-\frac{1}{4}\right] \left[\delta-\frac{3}{4}\right]\frac{1}{\sin^{2}u}
    -4\left[ \gamma-\frac{1}{4}\right] \left[\gamma-\frac{3}{4}\right]\frac{1}
      {\cos^{2}u}
  \Big\}\end{array}\psi=0. 
  \end{eqnarray*}
The substitutions
\letra
 \begin{eqnarray}\label{trig-1}
  z=\cos^2u,\qquad \psi(u)={\psi}[u(z)]=
  z^{\delta-\frac{1}{4}}(z-1)^{\gamma-\frac{1}{4}}U(z),
   \qquad\left[0\leq z\leq 1\right] 
   \end{eqnarray}
 transform the Schr\"odinger equation into a confluent Heun equation with 
 \begin{eqnarray}\label{trig-2}
 &\begin{array}{l}
 z_0=1,\qquad B_{1}=-2\delta, \qquad  B_{2}=2\gamma+2\delta, \qquad 
 B_{3}=-\frac{{\cal E}}{4}+\left(\gamma+\delta-\frac{1}{2}\right)^{2},\end{array} 
 \vspace{2mm}\nonumber\\
 &i\omega=\pm\beta,\qquad i\eta= \mp(\ell+\delta+\gamma),
 \end{eqnarray}
where the upper or the lower sign for ($i\eta,i\omega)$ must 
be taken throughout. Ushveridze supposed that $\ell=0,1,2,\cdots$ but, 
if $\gamma$ and $\delta$ are given by (\ref{4casos}),
 %
 %
the potential is quasisolvable even when $\ell$ 
 is a positive half-integer. These trigonometric Razavy potentials 
are ruled by Whittaker-Hill equations and will not be considered
in the following. We will see that it is convenient to express
the solutions as expansions in power series.

\subsection{Solutions in power series (Baber-Hass\'e)}

Finite-series solutions are obtained by taking $i\omega=\beta$ 
and $i\eta=-\ell-\gamma-\delta$
in $U_1^{\text{baber}}$ and $U_5^{\text{baber}}$. However, only 
 $U_5^{\text{baber}}$ satisfies the condition $\alpha_n\gamma_{n+1}>0$
 for real energies;
 the corresponding solution, denoted by $ \psi_{5}(u)$, 
is
 \antiletra\letra
 \begin{eqnarray}\label{barber-finita}
 \psi_{5}^{\text{baber}}(u)=e^{\beta \cos^2{u}}
 \left[\cos^2{u}\right]^{\delta-\frac{1}{4}}
 \left[1-\cos^2{u}\right]^{\gamma-\frac{1}{4}}
 \displaystyle \sum_{n=0}^{\ell}\begin{array}{l}(-)^na_{n}^{5}
 \left[\cos^2{u}\right]^{n},\;\;\left[\delta\geq\frac{1}{4},
 \;\gamma\geq\frac{1}{4}\right]\end{array}
 \end{eqnarray}
where recurrence relations for the
 coefficients are given by ($a_{-1}^{5}=0$)
   \begin{eqnarray}\label{B5}
   \begin{array}{l}
    \left(n+2\delta\right)
   \left(n+1\right)a_{n+1}^{5}+
   B_n^5\;a_{n}^{5}
  -2\beta\left(n-\ell-1\right)
   a_{n-1}^{5}=0,\quad \text{ where}\end{array}\nonumber\\
\begin{array}{l}
B_n^5=n\left(n+2\gamma+2\delta-1-2\beta\right)-\frac{{\cal E}}{4}+
   \left(\gamma+\delta-\frac{1}{2}\right)^{2}
   +2\beta(\ell+\gamma).
   \end{array}
   \end{eqnarray}
The conditions $\delta\geq1/4$ and $\gamma\geq1/4$ assure
that the solutions are bounded at $\cos{u}=0$ and $\cos{u}=\pm 1$,
respectively. 

Infinite-series solutions results from 
$i\omega=-\beta$ and $i\eta=\ell+\gamma+\delta$
in $U_1^{\text{baber}}$ and $U_5^{\text{baber}}$.   
Denoting these solutions by $\bar{\psi}_{1}$ and $\bar{\psi}_{5}$,
respectively, we have
\antiletra\letra
\begin{eqnarray}\label{barber-inf5}
\bar{\psi}_{1}^{\text{baber}}(u)=e^{-\beta \cos^2{u}}\left[\cos^2{u}\right]^{\delta-\frac{1}{4}}
\left[1-\cos^2{u}\right]^{\gamma-\frac{1}{4}}
\displaystyle \sum_{n=0}^{\infty}(-1)^n\,\bar{a}_{n}^{1}
\left[1-\cos^2{u}\right]^{n},
\end{eqnarray}
where the coefficients are given by ($\bar{a}_{-1}^{1}=0$)
\begin{eqnarray}\label{barB1}
&\begin{array}{l}
 \left(n+2\gamma\right)
\left(n+1\right)\bar{a}_{n+1}^{1}+
\bar{B}_n^1\;\bar{a}_{n}^{1}-2\beta\left(n+\ell+2\gamma+2\delta-1\right)
\bar{a}_{n-1}^{1}=0 \quad\text{ with}
\end{array}\nonumber\\
%
&%
\begin{array}{l}
\bar{B}_n^1= n\left(n+2\gamma+2\delta-1-2\beta\right)-
\frac{{\cal E}}{4}+\left(\gamma+\delta-\frac{1}{2}\right)^{2}
-2\beta\gamma\end{array}
\end{eqnarray}
and 
\antiletra\letra
\begin{eqnarray}\label{barber-inf5-b}
\bar{\psi}_{5}^{\text{baber}}(u)=e^{-\beta \cos^2{u}}
\left[\cos^2{u}\right]^{\delta-\frac{1}{4}}
\left[1-\cos^2{u}\right]^{\gamma-\frac{1}{4}}
\displaystyle \sum_{n=0}^{\infty}\bar{a}_{n}^{5}
\left[-\cos^2{u}\right]^{n}
\end{eqnarray}
satisfying the recurrence relations ($\bar{a}_{-1}^{5}=0$)
\begin{eqnarray}
&\begin{array}{l}\label{barB5}
\left(n+2\delta\right)
\left(n+1\right)\bar{a}_{n+1}^{5}+
\bar{B}_n^5\;\bar{a}_{n}^{5}+
2\beta\left(n+\ell+2\gamma+2\delta-1\right)
\bar{a}_{n-1}^{5}=0 \quad \text{with}\end{array}\nonumber\\
&\begin{array}{l}
\bar{B}_n^5= n\left(n+2\gamma+2\delta-1+2\beta\right)-
\frac{{\cal E}}{4}+\left(\gamma+\delta-\frac{1}{2}\right)^{2}
+
2\beta(\ell+\gamma+2\delta).
\end{array}
\end{eqnarray}

For $\delta\geq1/4$ and $\gamma\geq1/4$,
the solutions $\bar{\psi}_{1}$ and  $\bar{\psi}_{5}$
are both convergent and bounded for whole interval $0\leq \cos^2{u}\leq 1$.
In this manner, we have found only one expression for finite-series
solutions, but two expressions for infinite-series solutions,
 $\bar{\psi}_1$ and  $\bar{\psi}_5$. However, the latter
 are formally identical, as we will see.

In effect, by using the relation
\antiletra
\begin{eqnarray*}
(1+x)^n=\sum_{k=0}^{n}\left(\begin{array}{l}
n\\
k
\end{array}\right)
x^{k}=\sum_{k=0}^{n}\frac{n!\;x^n}{k!\;(n-k)!}
\end{eqnarray*}
and taking $z=\cos^2{u}$ in $\bar{\psi}_1$,we find
\begin{eqnarray*}
&&\sum_{n=0}^{\infty}(-)^n\,\bar{a}_{n}^{1}
\left[1-z\right]^{n}=
\sum_{n=0}^{\infty}(-)^n\,n!\;\bar{a}_{n}^{1}\sum_{k=0}^{n}
\frac{\left(-z\right)^{k}}{(n-k)!\,k!}=\sum_{n=0}^{\infty}
(-1)^n\,n!\;\bar{a}_{n}^{1}
\sum_{k=0}^{\infty}
\frac{\left(-z\right)^{k}}{(n-k)!\,k!}\nonumber\\
&&=\sum_{k=0}^{\infty}\frac{(-z)^k}{k!}\sum_{n=k}^{\infty}
\frac{\left(-\right)^{n}n!\;\bar{a}_{n}^{1}}{(n-k)!}=
\sum_{k=0}^{\infty}\frac{(-z)^k}{k!}\sum_{m=0}^{\infty}
\frac{\left(-\right)^{m+k}(m+k)!\;\bar{a}_{m+k}^{1}}{m!}\nonumber\\
&&
=\sum_{n=0}^{\infty}
(-z)^n\left[\frac{1}{n!}\sum_{m=0}^{\infty}
\frac{\left(-\right)^{m+n}(m+n)!}{m!}\;\bar{a}_{n+m}^{1}\right].
\end{eqnarray*}
Hence, both $\bar{\psi}_1$ and $\bar{\psi}_5$
are give by the same series of $(-z)^n$ provided that
\begin{eqnarray}\label{consitencia-1}
\bar{a}_{n}^{5}=\frac{(-)^n}{n!}\sum_{m=0}^{\infty}
\frac{\left(-\right)^{m}(m+n)!}{m!}\;\bar{a}_{n+m}^{1}.
\end{eqnarray}
Similarly, by starting from $\bar{\psi}_5$, we get
\begin{eqnarray*}
\sum_{n=0}^{\infty}\bar{a}_{n}^{5}
\left[-z\right]^{n}=
\sum_{n=0}^{\infty}
(-)^n[1-z]^n\left[\frac{1}{n!}\sum_{m=0}^{\infty}
\frac{\left(-\right)^{m+n}(m+n)!}{m!}\;\bar{a}_{n+m}^{5}\right]
\end{eqnarray*}
and, as a consequence, both $\bar{\psi}_1$ and $\bar{\psi}_5$
are are given by the same series of $(1-z)^n$ if
\begin{eqnarray}\label{consitencia-2}
\bar{a}_{n}^{1}=\frac{(-)^n}{n!}\sum_{l=0}^{\infty}
\frac{\left(-\right)^{l}(l+n)!}{l!}\;\bar{a}_{n+l}^{5}.
\end{eqnarray}
%
%
%
%
%
%

\subsection{Solutions in series of confluent hypergeometric functions}
Using the correspondences (\ref{hiper-baber}),
we replace the preceding Baber-Hass\'e  expansions
by series $\mathring{\textsf{U}}_i$ of regular confluent
hypergeometric functions as follows: 
\begin{eqnarray*}
&&\begin{array}{l}
U_5^{\text{baber}}\mapsto\mathring{\textsf{U}}_8 \text{ with } 
(i\omega,i\eta)=(\beta,-\ell-\gamma-\delta): \text{ finite series }
{\psi}_8,
\end{array}\\
%
&&\begin{array}{l}
\left(U_1^{\text{baber}},U_5^{\text{baber}}\right)\mapsto
\left(\mathring{\textsf{U}}_4,\mathring{\textsf{U}}_8\right) \text{ with } 
(i\omega,i\eta)=(-\beta,\ell+\gamma+\delta): \text{ infinite series }
\left({\bar{\psi}}_4,{\bar{\psi}}_8\right).
\end{array}\end{eqnarray*}
Thus, instead of (\ref{barber-finita}), we get the finite-series solutions
 \letra
 \begin{equation}\label{nossa1}
 \psi_{8}(u)=e^{\beta \cos^2{u}}
 \left[\cos^2{u}\right]^{\delta-\frac{1}{4}}
 \left[\sin^2{u}\right]^{\gamma-\frac{1}{4}}
 \displaystyle \sum_{n=0}^{\ell}\begin{array}{l}
 \mathring{b}_{n}^{8}
 \left[-2\beta\sin^2{u}\right]^{n}\tilde{\Phi}\left(n-\ell,
n+2\gamma;2\beta\sin^2u\right), 
\end{array}
 \end{equation}
where, in the recurrence relations 
(\ref{r1a}) for $\mathring{b}_{n}^{8}$,
   \begin{eqnarray}
   \begin{array}{l}
\mathring{\alpha}_n^8=2\beta
   \left(n+1\right),\qquad  \mathring{\beta}_n^8=B_n^5\text{ of }\text{Eq. (\ref{B5})},\qquad
 \mathring{\gamma}_n^8= -\left(n-\ell-1\right)(n-1+2\delta).\end{array}
   \end{eqnarray}
For infinite series, we obtain
\antiletra
\letra
 \begin{eqnarray}\label{nossa2}
 \bar\psi_{4}(u)&=&e^{-\beta \cos^2{u}}
 \left[\cos^2{u}\right]^{\delta-\frac{1}{4}}
 \left[\sin^2{u}\right]^{\gamma-\frac{1}{4}}\nonumber\\
 &\times&
 \displaystyle \sum_{n=0}^{\infty}\begin{array}{l}
 \bar{\mathring{b}}_{n}^{4}
 \left[-2\beta\cos^2{u}\right]^{n}\tilde{\Phi}\left(n+\ell+2\gamma+2\delta,
n+2\delta;2\beta\cos^2u\right), 
\end{array}
 \end{eqnarray}
where, in the recurrence relations 
(\ref{r1a}) for $\bar{\mathring{b}}_{n}^{4}$,
   \begin{equation}
   \begin{array}{l}
\mathring{\alpha}_n^4=2\beta
   \left(n+1\right),\quad \mathring{\beta}_n^4=\bar{B}_n^1\text{ of }\text{Eq. (\ref{barB1})},\quad
 \mathring{\gamma}_n^4= -\left(n+\ell-1+2\gamma+2\delta\right)(n-1+2\gamma),\end{array}
   \end{equation}
and
\antiletra
\letra
 \begin{eqnarray}\label{nossa3}
 \bar\psi_{8}(u)&=&e^{-\beta \cos^2{u}}
 \left[\cos^2{u}\right]^{\delta-\frac{1}{4}}
 \left[\sin^2{u}\right]^{\gamma-\frac{1}{4}}\nonumber\\
&\times&\displaystyle \sum_{n=0}^{\infty}\begin{array}{l}
 \bar{\mathring{b}}_{n}^{8}
 \left[2\beta\sin^2{u}\right]^{n}\tilde{\Phi}\left(n+\ell+2\gamma+2\delta,
n+2\gamma;-2\beta\sin^2u\right), 
\end{array}
 \end{eqnarray}
where, in the recurrence relations 
(\ref{r1a}) for $\bar{\mathring{b}}_{n}^{8}$,
   \begin{equation}
   \begin{array}{l}
\mathring{\alpha}_n^8=-2\beta
   \left(n+1\right),\quad  \mathring{\beta}_n^8=\bar{B}_n^5
   \text{ of }\text{Eq. (\ref{barB5})},
   \quad
 \mathring{\gamma}_n^8= -\left[n+\ell-1+2\gamma+2\delta\right][n-1+2\delta].\end{array}
   \end{equation}
The solutions $\bar{\psi}_{4}$ and  $\bar{\psi}_{8}$
are both convergent and bounded for whole interval $0\leq \cos^2{u}\leq 1$.
In this manner, we have one expression for finite-series
solutions, but two expressions for infinite-series solutions,
 $\bar{\psi}_4$ and  $\bar{\psi}_8$. Up to now we have not been able to show that 
  $\bar{\psi}_4$ and  $\bar{\psi}_8$ are linearly dependent. 
In any case, the power series solutions 
(\ref{barber-inf5}) and (\ref{barber-inf5-b}) do not present this ambiguity.

\section{Conclusion}

{We have found new solutions for the CHE and applied them to the Schr\"odinger 
equation with quasiexactly solvable potentials. 
To get suitable infinite-series solutions for 
the hyperbolic potential (\ref{potencial}) we have used two-sided series.
For the hyperbolic Razavy-type potentials we have found
only even infinite-series solutions convergent and bounded for
any value of the independent variable; to get odd solutions
it is necessary to use two solutions with different domains of convergence. 
Excluding the Razavy-type potentials, we have seen that it is
advantageous to use the expansions in series of Coulomb
wave functions to get infinite-series solutions appropriate
for $2/4<\delta\leq3/4$.}

{For the trigonometric potential (\ref{ush-trigonometrico}), we can
use one-sided series to obtain bounded infinite-series solutions.
In fact, the Schr\"{o}dinger equation 
may be solved by using the Baber-Hass\'e expansions in power series
provided that we take the minimal solutions of the recurrence relations 
for the series coefficients }(then, the series converge for any finite value of $z$ \cite{baber}). We can also
use expansions in series of regular hypergeometric functions but, in this case,
it is necessary additional study to decide on the duplicity of
solutions referred to in section V.B. This issue occurs also for the trigonometric 
Razavy-type potentials which are given by expression (\ref{ush-trigonometrico}) 
along with the restrictions (\ref{4casos}): {solutions for 
this problem need further considerations. }
%



{Another question consists in knowing if the one-sided expansions 
$\mathring{U}_i^{\infty}$ in series
of irregular confluent hypergeometric functions can 
play the role of the Hylleraas \cite{hylleraas} and Jaff\'e \cite{jaffe} 
solutions which so far have been used to represent bound states of hydrogen 
moleculelike ions. While it is difficult to determine the
conditions for the convergence of the 
Hylleraas and Jaff\'e solutions \cite{leaver}, 
the conditions (\ref{regiao-1}) for the convergence of $\mathring{U}_i^{\infty}$ 
follow from the Raabe test.}

{On the other side, in section III we have seen that the expansions $U_i^{\infty}$ 
and $U_i$ of a given set of two-sided solutions 
may cover the entire complex plane $z$, contrary to the solutions 
in series of Coulomb wave functions. This suggests that such
solutions could be useful for some astrophysical problems.
In effect, in order to solve the
radial Teukolsky equations,  Otchik \cite{otchik1} proposed to
match an expansion in series of hypergeometric
functions and another in series of Coulomb functions, both
having the same series coefficients but 
converging in different regions (see also \cite{mano1,mano2}).
The use of $U_i^{\infty}$ and $U_i$ would give, by means of (\ref{continuacao}),
a solution ${\textsf{U}}_i$ which is a linear combination of $U_i^{\infty}$ and $U_i$.}

At last we mention that the new solutions for the CHE (\ref{gswe}) can afford 
solutions for an equation 
called reduced CHE \cite{kazakov-1}. We write the latter in the form
\antiletra
\begin{equation}\label{incegswe}
z(z-z_{0})\frac{d^{2}U}{dz^{2}}+(B_{1}+B_{2}z)
\frac{dU}{dz}+
\left[B_{3}+q(z-z_{0})\right]U=0,\qquad\left[q\neq0\right]
\end{equation}
where $z_0$, $B_{i}$ and $q$ ($q\neq0$) are constants. 
In Eqs. (\ref{gswe}) and (\ref{incegswe}),
if $z_0\neq 0$ then $z=0$ and $z=z_{0}$ are regular singular
points with exponents ($0,1+B_{1}/z_{0}$)
and ($0,1-B_{2}-B_{1}/z_{0}$), respectively.
However, at the irregular singular point $z=\infty$ the expected behavior of the 
solutions are \cite{eu}
\begin{eqnarray*}\label{thome1}
U(z)\sim e^{\pm i\omega z}\ z^{\mp i\eta-(B_{2}/2)}\ 
\quad \text {for Eq.    (\ref{gswe}) },
\qquad
U(z)\sim
e^{\pm 2i\sqrt{qz}}\ z^{(1/4)-(B_{2}/2)} \quad \text{for 
Eq. (\ref{incegswe})}.
\end{eqnarray*}
{Eq. (\ref{incegswe}) is obtained
by applying the so-called Whittaker-Ince limit \cite{lea-1,eu}}
\begin{eqnarray}\label{ince}
\omega\rightarrow 0, \quad
\eta\rightarrow
\infty \quad \mbox{such that }\quad \ 2\eta \omega =-q,
\qquad [\text{Whittaker-Ince limit}]
\end{eqnarray}
to equation (\ref{gswe}). 
By accomplishing the above limit in the same manner we have done with regard to the expansions
in Coulomb wave functions \cite{lea-2013}, we will find that the solutions
in series of confluent hypergeometric functions give new expansions in series 
of Bessel functions for the reduced CHE. In particular, we will get new solutions in series
of Bessel functions also for the Mathieu equation by the reason that this
is a special case of the reduced CHE \cite{lea-2013}.

\appendix

\section{Recurrence relations in matrix form}
\protect\label{A}
\setcounter{equation}{0}
\renewcommand{\theequation}{A.\arabic{equation}}

%


%
%
The three-term recurrence relations 
having the form (\ref{r1a}) can be written as
\begin{eqnarray}\label{matriz-tridiagonal}
\left[
\begin{array}{lcccl|llr}
\mathring\beta_{0} & \mathring\alpha_{0} & 0 & &� &� &\\
\mathring\gamma_{1}&\mathring\beta_{1} & \mathring\alpha_{1} & & �&
\\
0�&\mathring\gamma_{2} & \mathring\beta_{2}&
\mathring\alpha_{2}& &� \vspace{1mm}\\
& & \ddots �&\ddots& \ddots &�& \vspace{1mm}\\
&� & �&\mathring\gamma_{\text{\tiny N}} & \;\mathring\beta_{\text{\tiny N}}&
\mathring\alpha_{\text{\tiny N}}\\
\hline
 & & �& & \mathring\gamma_{\text{\tiny N}+1}&
\mathring\beta_{\text{\tiny N}+1}& \mathring\alpha_{\text{\tiny N}+1} \\
 & &� & �& & \mathring\gamma_{\text{\tiny N}+2}&
\mathring\beta_{\text{\tiny N}+2}& \mathring\alpha_{\text{\tiny N}+2}\\
 & & & & &\; \; \; \;\;\ddots  & \;\;\;\ddots 
& \ddots
\end{array}
\right]
\left[\begin{array}{l}
\mathring{b}_{0} �\\
\mathring{b}_{1} \\
\mathring{b}_{2} \\
\vdots\\
%
%
\mathring{b}_{\text{\tiny N}}\\
\hline
\mathring{b}_{\text{\tiny N}+1}\\
\mathring{b}_{\text{\tiny N}+2}\\
\vdots
\end{array}
\right]=
\left[\begin{array}{c}
0 \\
0 \\
0 \\
\vdots\\
0\\
\hline
0\\
0\\
\vdots
\end{array}
\right],
\end{eqnarray}
where we have suppressed the upper indices and split
the matrix into blocks. 

The previous relations hold only if the series begin at $n=0$. 
%
In fact, in some cases the series truncate on 
the left-hand side and, then,
the series begin at $n>0$ (see
page 171 of \cite{arscott}, Ex. 2); in other cases 
cases the series truncate on 
the right-hand side leading to finite series 
(p. 146 of \cite{arscott}). 
Specifically, for one-sided series 
solutions, 
%
\begin{itemize}
\itemsep-3pt
\item
If $\mathring{\alpha}_{n=\text{\tiny N}}^i=0$ for some 
$\text{\small N}\geq 0$, then
the series begins 
at $n=\text{\small N}+1$, that is, $\mathring{b}_{0}^i=\cdots=
\mathring{b}_{\text{\tiny N}}^i=0$. In this case,
only the right lower block of the matrix
is relevant. If we set $n= m+N$ and relabel 
the series coefficients, we 
reobtain series beginning at $m=0$ with
recurrence relations like (\ref{r1a}).
\item
If $\mathring{\gamma}_{n=\text{\tiny N}+1}^i=0$ 
for some $\text{\small N}\geq 0$, then the series
terminates at $n=\text{\small N}$ ($0\leq n\leq \small N$), that is, 
$\mathring{b}_{\text{\tiny  N}}^i=\mathring{b}_{\text{\tiny N}+1}^i=\cdots=0$.
In this case, only the left upper block of the matrix
is relevant.
%
\end{itemize}

 On the other hand, for two-sided series solutions 
 we rewrite the relations (\ref{recursion-1}) as
  \begin{eqnarray}\label{matriz-tridiagonal-nu}
  \left[
  \begin{array}{lllllll}
  . & \ . & \ . && & \vspace{1mm}\\
          & \gamma_{n}^{i} & \beta_{n}^{i} &
  \alpha_{n}^{i} & \vspace{1mm}\\
  %
  %
       &  & \gamma_{n+1}^{i} &
  \beta_{n+1}^{i} & \alpha_{n+1}^{i} & \vspace{1mm}\\
    &  &  & 
    \gamma_{n+2}^{i} &
  \beta_{n+2}^{i} & \alpha_{n+2}^{i} & \vspace{1mm}\\
     &   & &  &  \ .
  &  \ .&   \ . 
  %
  \end{array}
  \right]
  \left[\begin{array}{l}
  \ .\vspace{1mm}\\
  %
  %
  b_{n-1}^{i}\vspace{1mm}\\
  %
  b_{n}^{i}\vspace{1mm}\\
  b_{n+1}^{i}\vspace{1mm}\\
  %
  %
  \ .
  \end{array}
  \right]=\mathbf{0}
  \qquad [-\infty<n<\infty]
  \end{eqnarray}
  where $\mathbf{0}$ denotes the null column vector. 
  In this case, $\alpha_n^i$, $\beta_n^i$ and $\gamma_n^i$
  depend on a parameter $\nu_i$. By determining $\nu_i$ 
    such that $\alpha_{-1}^i=0$, we obtain one-sided series 
    expansions ($n\geq 0$) out of two-sided series expansions: 
for the expansions in series of hypergeometric functions 
considered in section III, we find $\nu_i=0$.

\section{Remarks on the spheroidal equation}
\protect\label{B}
\setcounter{equation}{0}
\renewcommand{\theequation}{B.\arabic{equation}}
The (ordinary) spheroidal wave 
equation \cite{nist},
\begin{equation}\label{esferoidal}
\begin{array}{l}
\frac{d}{dx}\left[\left( 1-x^2\right) \frac{dX(x)}{dx} \right]+
\left[ \gamma^2(1-x^2)+\bar{\lambda}-
\frac{\mu^2}{1-\
x^2}\right] X(x)=0,
\end{array}
\end{equation}
is a particular case of the CHE. In effect, 
%
%
%
%
the substitutions
 %
 \begin{eqnarray}\label{esferoidal-2}
 x=1-2z,\qquad X(x)=z^{{\mu}/{2}
 }\ (z-1)^{{\mu}/{2}}U(z),
 \end{eqnarray}
transform equation (\ref{esferoidal}) into
 \begin{eqnarray*}
 z(z-1)\frac{d^{2}U}{dz^{2}}+
 \left[
 -\left(\mu+1\right)+2
 \left(\mu+1\right)z
 \right]
 \frac{dU}{dz}+
 \left[
 \mu\left(
 \mu+1\right)-\bar{\lambda}+4\gamma^{2}z(z-1)
 \right]U=0,
 \end{eqnarray*}
 which is a CHE (\ref{gswe}) with
 \begin{eqnarray}\label{esferoidal-3}
 z_0=1,\qquad B_2=-2B_1=2(\mu+1),
 \quad B_3=\mu(\mu+1)-\bar{\lambda},\qquad
  \eta=0,
 \qquad \omega^2=4\gamma^2.
 \end{eqnarray}
 Therefore, the spheroidal equation is 
 a CHE with $z_0=1$, $\eta=0$ and $B_2=-2B_1$,
 %
that is,
%
\begin{eqnarray}
\displaystyle z(z-1)\frac{d^{2}U}{dz^{2}}+(B_{1}-2B_{1}z)\frac{dU}{dz}
+\displaystyle\left[B_{3}+\omega^{2}z(z-1)\right]U=0
\quad(\text{spheroidal equation})
\end{eqnarray}

For some solutions of the spheroidal equation, 
the definition (\ref{phi-tilde}) for
$\tilde{\Phi}(a,c;y)$ becomes inappropriate because 
$c-a$ in $\Gamma(c-a)$ is zero or a negative integer. To avoid
this problem, first we rewrite the solutions in terms of
$\Phi(a,c;y)$ by putting
\begin{eqnarray}\label{esf-2casos}
\begin{array}{l}
\mathring{b}_n^i\tilde{\Phi}(a,c;y)=\frac{ \tilde{b}_n^i}{\Gamma(c)}{\Phi}(a,c;y)
\;\text{ for }\: \mathring{U}_1(z),\;\mathring{U}_3(z),\;
\mathring{U}_5(z),\;\mathring{U}_7(z);\vspace{2mm}\\
\mathring{b}_n^i\tilde{\Phi}(\bar{a},\bar{c};y)=  \bar{b}_n^i 
{\Phi}(\bar{a},\bar{c};y)
\;\text{ for }\: \mathring{\textsf{U}}_2(z),\;\mathring{\textsf{U}}_4(z),\;
\mathring{\textsf{U}}_6(z),\;\mathring{\textsf{U}}_8(z),
\end{array}
\end{eqnarray}
where the equations for $ \tilde{b}_n^i $ and $ \bar{b}_n^i $
are obtained from the ones for $ \mathring{b}_n^i$ 
by taking 
\[ \tilde{b}_n^i=\Gamma(c-a) \mathring{b}_n^i,\qquad 
\bar{b}_n^i={\Gamma(\bar{c}-\bar{a})} \mathring{b}_{n}^i\big/{\Gamma(\bar{c})}.\]
Hence, the forms of the recurrence relations, similar to (\ref{r1a}),
are ($\tilde{b}_{-1}^i=\bar{b}_{-1}^i=0$)
\begin{eqnarray}\label{b-tilde}
\tilde{\alpha}_{n}^i\  \tilde{b}_{n+1}^i+\tilde{\beta}_{n}^i\ 
\tilde{b}_{n}^i+
\tilde{\gamma}_{n}^i\  \tilde{b}_{n-1}^i=0,
\end{eqnarray}
%
%
\begin{eqnarray}\label{b-bar}
\bar{\alpha}_{n}^i\  \bar{b}_{n+1}^i+\bar{\beta}_{n}^i\ 
\bar{b}_{n}^i+
\bar{\gamma}_{n}^i\  \bar{b}_{n-1}^i=0.
\end{eqnarray}
We find that the solutions $\mathring{\textsf{U}}_i(z)$
of the second case given in (\ref{esf-2casos})  
reduce to power-series expansions since 
$\Phi(\bar{a},\bar{c};y)= \Phi(\bar{a},\bar{a};y)$$=$$\exp{(y)}$,
while the solutions $\mathring{U}_i^ {\infty}(z)$ 
and $\mathring{U}_i(z)$ may be expressed as 
expansions in series of incomplete 
gamma functions, $\Gamma(\beta,y)$ and $\gamma(\beta,y)$,
by means of \cite{nist}
\begin{eqnarray*}\label{gammas}
\Gamma(\beta,y)=e^{-y} y^{\beta}\Psi(1,1+\beta;y),\qquad
\gamma(\beta,y)=e^{-y} y^{\beta} \Phi(1,1+\beta;y)/\beta.
\end{eqnarray*}
The function $\gamma(\beta,y)$ is valid only if 
$\beta$ is not zero or negative integer: in this case
\[\Gamma(\beta,y)+\gamma(\beta,y)=\Gamma(\beta).\]

As examples, below we write two sets of solutions.  The 
first one is
%
%
%
\begin{equation}\label{che-primeiro-set-esf}
\begin{array}{l}
\left[\begin{array}{l}
\mathring{U}_1^{\infty}\vspace{2mm}\\
\mathring{U}_1
\end{array}\right]=\begin{array}{l}e^{i\omega z}z^{1+{B_1}}
[z-1]^{1+B_1}\end{array}
\displaystyle\sum_{n=0}^{\infty}
\left[\begin{array}{l}
\mathring{b}_n^{1}\Psi\left(2+B_1,2+{B_1}-n; - 2i\omega z\right)
\vspace{2mm}\\
\frac{\tilde{b}_n^{1}}{\Gamma(2+B_1-n)}\Phi\left(2+B_1,2+{B_1}-n;
 - 2i\omega z\right)
\end{array}\right],
\vspace{2mm}\\
\mathring{\textsf{U}}_1(z) = 
\begin{array}{l}
e^{i\omega z}
[z-1]^{1+B_1}\end{array}
\displaystyle\sum_{n=0}^{\infty}\begin{array}{l}
\mathring{b}_{n}^{1}\left(2i\omega z\right)^n \tilde{\Phi}
\left(n+1,
n-{B_1};-2i\omega z\right),\end{array}
\end{array}
\end{equation}
where the coefficients $\mathring{b}_n^1$ satisfy the recurrence relations 
(\ref{r1a}) with
\begin{eqnarray}\label{alfa-esf}
&\begin{array}{l}
\mathring{\alpha}_n^1=- 2i\omega (n+1), \qquad
\mathring{\beta}_n^1=n\left[n+1+
2i\omega \right]+
i\omega \left[2+B_1\right]-B_1[1+B_1]+B_3,\end{array}
\nonumber\\
&\begin{array}{l}
\mathring{\gamma}_n^1=-n
\left[n+1+B_1\right],\end{array}
\end{eqnarray}
whereas the $\tilde{b}_n^1$ satisfy (\ref{b-tilde}) with
%
\begin{eqnarray}
\begin{array}{l}
\tilde{\alpha}_n^1= 2i\omega (n+1)^2, \qquad
\tilde{\beta}_n^1=\mathring{\beta}_n^1,\qquad 
%
\tilde{\gamma}_n^1=
n+1+B_1.\end{array}
\end{eqnarray}
The second set takes the form
%
\begin{equation}
\begin{array}{l}
\left[\begin{array}{l}
\mathring{U}^{\infty}_2(z)\vspace{2mm}\\
\mathring{U}_2(z)
\end{array}\right]=\begin{array}{l}e^{i\omega z}[z-1]^{1+B_1}\end{array}
\displaystyle\sum_{n=0}^{\infty}
\mathring{b}_n^{2}
\left[\begin{array}{l}
\Psi\left(1,-{B_1}-n; - 2i\omega z\right)
\vspace{2mm}\\
\tilde{\Phi}\left(1,-{B_1}-n; - 2i\omega z\right)
\end{array}\right],
\vspace{2mm}\\
\mathring{\textsf{U}}_2(z) = 
\begin{array}{l}
e^{-i\omega z}z^{1+{B_1}}
[z-1]^{1+B_1}\end{array}
\displaystyle\sum_{n=0}^{\infty}\begin{array}{l}
\bar{b}_{n}^{2}\left[2i\omega z\right]^n,\end{array}
\end{array}
\end{equation}
where the $\mathring{b}_n^2$ satisfy the recurrence relations 
(\ref{r1a}) with
\begin{eqnarray}
&\begin{array}{l}
\mathring{\alpha}_n^2=-2i\omega (n+1), \qquad
\mathring{\beta}_n^2=n\left[n+3+2B_1+2i\omega\right]+
 i \omega \left[2+B_1\right]+2+2B_1+B_3, 
\end{array}
\nonumber\\
&\begin{array}{l}
\gamma_n^2=-\left[n+1+B_1\right]^2,\end{array}
\end{eqnarray}
and $\bar{b}_n^2$ satisfy (\ref{b-bar}) with 
%
%
\begin{eqnarray}
&\begin{array}{l}
\bar{\alpha}_n^2=-2i\omega (n+1)\left[n+2+B_1\right], \qquad
\bar{\beta}_n^2=\mathring{\beta}_n^2, \qquad
\bar{\gamma}_n^2=-\left[n+1+B_1\right].\end{array}
\end{eqnarray}
\end{document}